# Radio-only and Radio-to-far-ultraviolet Spectral Energy Distribution Modeling of 14 ULIRGs: Insights into the Global Properties of Infrared Bright Galaxies

Subhrata Dey[1] , Arti Goyal[1] , Katarzyna Małek[2] , and Tanio Díaz-Santos[3,4]
[1] Astronomical Observatory of the Jagiellonian University, Orla 171, 30-244 Kraków, Poland
[2] National Centre for Nuclear Research, ul. Pasteura 7, 02-093 Warsaw, Poland
[3] Institute of Astrophysics, Foundation for Research and Technology-Hellas, GR-71110, Heraklion, Greece
[4] School of Sciences, European University Cyprus, Diogenes Street, Engomi, 1516 Nicosia, Cyprus

## Abstract

We present detailed spectral energy distribution (SED) modeling of 14 local ultraluminous infrared galaxies (ULIRGs) with outstanding photometric data from the literature covering the ultraviolet–infrared (FIR) and radio bands (∼50 MHz to ∼30 GHz). We employ the CIGALE SED fitting code to model the ultraviolet–FIR–radio SED. For the radio-only SED modeling, we use the UltraNest package, leveraging its nested sampling algorithm. Combining the results from our previous study on 11 luminous infrared galaxies (LIRGs), we discuss the global astrophysical properties of a sample of 25 starburst galaxies ($z < 0.5$). Their radio spectra are frequently characterized by bends and turnovers, with no indication of ULIRGs exhibiting more complicated SEDs than LIRGs despite showing more signs of interactions. Including radio measurements in the CIGALE modeling constrained the dust luminosity and star formation rate (SFR) estimates by more than 1 order of magnitude better than previously reported for starburst galaxies. We show that total and nonthermal radio luminosity at 1.4 and 4.8 GHz frequencies can be good estimators of recent SFRs for all LIRGs and those ULIRGS with an insignificant influence of active galactic nuclei. A weaker but still significant correlation is observed between radio SFRs at 1.4 GHz and old (averaged over 100 Myr) SFRs based on SED modeling, indicative of multiple episodes of starburst activity during their lifetime. The thermal radio luminosity at 4.8 GHz is a better tracer of recent star formation than the thermal luminosity at 1.4 GHz. Statistically, our modeled nonthermal radio spectral indices do not significantly correlate with redshift, stellar mass, SFR, specific SFR, and dust mass.

*Unified Astronomy Thesaurus concepts:* Radio continuum emission (1340); Ultraluminous infrared galaxies (1735); Spectral energy distribution (2129); Galaxy photometry (611); Interstellar medium (847)

*Supporting material:* figure set

## 1. Introduction

Luminous infrared galaxies (LIRGs; $L_{\rm IR}$[8–1000 $\mu$m] $\lesssim 10^{11-12} L_\odot$, where $L_\odot$ is the solar luminosity) and ultraluminous infrared galaxies (ULIRGs; $L_{\rm IR}$[8–1000 $\mu$m] $\gtrsim 10^{12} L_\odot$) are excellent laboratories for studying star formation and active galactic nuclei (AGN; Sanders et al. 1988). ULIRGs could be linked with the hierarchical transformation of star-forming galaxies into elliptical galaxies, the formation of quasars, and the coevolution of supermassive black holes (SMBH) and stellar bulges (Sanders et al. 1988; Hopkins et al. 2006). The high infrared (IR) luminosities in ULIRGs are associated with interactions and mergers between gas-rich galaxies that effectively distort their morphologies. These interactions and mergers funnel gas and material toward the galactic nuclei, which fuels the central starbursts and facilitates the accretion of matter onto an SMBH, leading to the growth of stellar and SMBH masses (for a recent review; U 2022). The high-IR luminosity is a result of the absorption of ultraviolet (UV) emissions generated by enhanced star formation, either by dust (Rigopoulou et al. 1999; Lonsdale et al. 2006) or by the optically thick torus surrounding the AGN. Subsequently, this absorbed energy is reradiated in the IR wavelength range. The relative contribution of these two processes and their interconnection has been extensively studied (Armus et al. 2007; Spoon et al. 2007; Veilleux et al. 2009).

The spectral energy distribution (SED) of galaxies is shaped by physical processes occurring within the galaxy with imprints of its components such as stars, dust, AGN, etc. Although different emission mechanisms contribute to different wavelength regimes, a considerable interplay between galaxy components impacts the overall shape of the SED. The far-ultraviolet (FUV)–IR SEDs of galaxies probe stellar mass ($M_\star$) and star formation rates (SFRs) as the emission in this regime is dominated by stellar processes related to short-lived massive OB stars (∼8 $M_\odot$) and low-mass stars (Taylor et al. 2011; Kennicutt & Evans 2012). The computation of SFR is further complicated as the intrinsic SEDs are modified by dust, which absorbs the UV-optical photons and reemits them at IR wave bands (Salim & Narayanan 2020). The presence of AGN adds to this complexity, as its SED also spans the entire electromagnetic spectrum, from radio to X-rays to $\gamma$-rays (Blandford et al. 2019). AGN have prominent features such as the *IR bump* at 10–20 $\mu$m and an upturn in optical-UV, the *big blue bump* (Elvis et al. 2012; Krawczyk et al. 2013). The former is the reprocessed dust emission from the torus of the UV photons, while the latter originates from the SMBH's accretion disk. Studying the properties of the host galaxies in the presence of AGN is challenging because the central nuclei often outshine them, making it hard to analyze stellar populations. In obscured AGN, the host galaxy's light dominates the optical/near-IR SED, complicating estimates of intrinsic nuclear power. To effectively address this issue, a widely used strategy involves fitting the SED by utilizing diverse combinations of theoretical models and empirical templates for each specific emission component (e.g., as demonstrated in da Cunha et al. 2008).

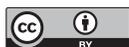

Radio SEDs probe processes related to star formation as well as the presence of AGN. The radio SED of galaxies is composed of steep spectrum synchrotron emission ($S \propto \nu^{-\alpha}$, where $\alpha = 0.75 \pm 0.1$ is the slope; Condon 1992) and flat spectrum thermal free–free emission (FFE; $S \propto \nu^{-0.1}$) processes (Condon 1992). Moreover, the shape of radio SED and its slope are used to probe the radiation laws and the heating/cooling mechanism of the interstellar medium (ISM) and AGN activity (e.g., Heesen et al. 2022). The radio SEDs of galaxies are further complicated by the presence of radio AGN, where the radio emission can be due to relativistic jets, or it can be from various sources, including star formation, AGN-driven winds, photoionized gas, low-power jets, and inner accretion disk coronal activity (Panessa et al. 2019). In recent times, it has become increasingly common to consider all components while modeling UV–IR-radio SEDs of galaxies, e.g., GRAZIL (Vega et al. 2008), CIGALE (Noll et al. 2009; Boquien et al. 2019), and PROSPECT (Thorne et al. 2023), albeit with simple assumptions such as the single power-law (PL) form of radio SEDs in CIGALE and PROSPECT (Yang et al. 2022; Thorne et al. 2023).

Since the radio luminosity results from two distinct emission mechanisms, we must decompose it into its respective components (synchrotron and FFE) in order to link it to star formation or AGN activity. The synchrotron emission results from an interaction of charged particles in a magnetic field that is accelerated in shocks produced by supernova remnants, and therefore, traces the supernova rate. The synchrotron emission radio component measures the SFR over timescales of $\sim$10–100 Myr for an assumed supernova rate and the initial mass function (IMF; Condon et al. 2002). The thermal FFE component is produced by the ionization of H II regions by the UV photons emitted from the same massive stars that eventually produce the supernova associated with the nonthermal synchrotron emission. The thermal FFE is a direct near-instantaneous measure of SFR. However, measuring thermal FFE in radio SEDs is complicated by the fact that low-frequency SEDs are dominated by nonthermal synchrotron emission ($\nu \lesssim 30$ GHz; Galvin et al. 2018). Furthermore, radio SEDs are further modified by free–free absorption (FFA) or synchrotron self-absorption (SSA) processes causing bends and turnovers (Clemens et al. 2010; Chyży et al. 2018; Galvin et al. 2018). Therefore, disentangling thermal emission from the total radio continuum requires model fitting using densely sampled SEDs and sophisticated modeling techniques (Clemens et al. 2008, 2010; Galvin et al. 2016, 2018; Dey et al. 2022) or high-frequency ($\geqslant 20$ GHz) radio measurements.

In our previous paper on this subject, Dey et al. (2022), we made the first attempt to simultaneously model the UV–IR–radio SEDs using CIGALE and the densely sampled radio SEDs ($\sim$70 MHz–15 GHz) for a sample of 11 LIRGs. Due to the inclusion of radio data in the SED modeling using CIGALE, we could estimate the dust luminosities with 1 order of magnitude better accuracy than previously reported (median uncertainties $\sim$0.03 dex compared to $\sim$0.3 dex in literature). We separated the nonthermal synchrotron and thermal FFE components in our radio SED modeling and obtained the radio nonthermal and thermal and total (combining the nonthermal and thermal emission) SFRs at 1.4 GHz using the radio luminosity scaling relations of Murphy et al. (2011). We then compared the nonthermal, thermal and total SFRs at 1.4 GHz with IR SFRs ($SFR_{IR}$) averaged over the last 10 and 100 Myr from the CIGALE analysis. We found a good correspondence between the $SFR_{IR}$ averaged over the last 10 Myr with the total and nonthermal radio SFRs, respectively. This indicates that the total nonthermal luminosities at 1.4 GHz are good indicators of recent star formation in LIRGs (on timescales of $\sim$10 Myr).

This work extends our approach to simultaneously modeling UV–IR–radio SEDs using CIGALE and densely sampled radio SEDs ($\sim$54 MHz–30 GHz) of a sample of 14 ULIRGs. We aim to investigate how high star formation activity impacts the galaxy properties by extending the sample of our LIRGs (11 objects; Dey et al. 2022). We note that many authors (e.g., da Cunha et al. 2010, Małek et al. 2017; Paspaliaris et al. 2021; Sokol et al. 2023) have previously studied UV–IR SEDs of large samples of ULIRGs, but only a few have modeled their radio SEDs in detail (with the exception of Clemens et al. 2010; Galvin et al. 2018; Nandi et al. 2021), due to the lack of radio data. Therefore, we restrict our sample to ULIRGs for which densely sampled radio SEDs and UV–IR SEDs are available. This is done to facilitate the complex modeling of the radio spectra with physically motivated models characterized by multiple free parameters. Section 2 provides the selection criteria for our targets, which are taken from Clemens et al. (2010), Galvin et al. (2018), and Nandi et al. (2021). The study presented here improves on previous modeling efforts in the following manner: (1) better coverage of the radio spectra, in particular, the inclusion of low-frequency flux measurements ($\sim$50–100 MHz), (2) comparison of physically motivated scenarios (simple and complex models with multiple free parameters) and using Bayesian inference to select the best-fit model and estimate the model parameters and their uncertainties.

## 2. Sample Selection

Our sample consists of 14 ULIRGs compiled from the following three studies:

1. Six galaxies from the sample of 31 galaxies from Clemens et al. (2010). We selected these targets as the rest of them did not have enough data coverage to perform the SED modeling using complex models with many free parameters.
2. Three galaxies from the sample of 12 galaxies from Galvin et al. (2016) as they did not perform radio SED modeling of these galaxies.
3. Five galaxies from the sample of 13 galaxies studied by Nandi et al. (2021). We selected these targets because we could not find enough data for the rest of them to perform the complex SED modeling with many free parameters.

These galaxies were selected on the basis of the availability of photometric data with a wide range of wavelength coverage from the UV–far-IR (FIR) bands accompanied by two decades of spectral coverage in the radio band. Our sample has a redshift range of 0.03–0.4285 and $\log_{10} L_{IR} \gtrsim 11.90$ $L_{\odot}$, justifying a ULIRG classification. The basic properties of our sample of ULIRGs, including dust luminosity from the literature, are listed in Table 1.

Table 1
Basic Information on Our Sample of ULIRGs

| Source Name | R.A. (J2000) | Decl. (J2000) | $z$ | $\log_{10}(L_{IR})$ ($L_\odot$) | Ref. |
|---|---|---|---|---|---|
| (1) | (2) | (3) | (4) | (5) | (6) |
| IRAS F00183-7111 | 00$^h$20$^m$34.$^s$69 | −70°55′26.″7 | 0.3300 | 12.93[a] | (1) |
| IRAS F03538-6432 | 03$^h$54$^m$25.$^s$21 | −64°23′44.″7 | 0.3007 | 12.62[b] | (1) |
| IRAS 08572+3915 | 09$^h$00$^m$25.$^s$39 | +39°03′54.″40 | 0.0582 | 12.09[c] | (2) |
| UGC 05101 | 09$^h$35$^m$51.$^s$60 | +61°21′11.″5 | 0.0393 | 11.90[c] | (2) |
| IRAS 10565+2448 | 10$^h$59$^m$18.$^s$12 | +24°32′34.″46 | 0.0431 | 11.98[c] | (2) |
| IRAS 12112+0305 | 12$^h$13$^m$46.$^s$00 | +02°48′38.″0 | 0.0733 | 12.28[d] | (3) |
| UGC 08058 | 12$^h$56$^m$14.$^s$23 | +56°52′25.″2 | 0.0421 | 12.49[c] | (2) |
| IRAS 13305-1739 | 13$^h$33$^m$16.$^s$54 | −17°55′10.″7 | 0.1483 | 12.21[d] | (3) |
| UGC 8696 | 13$^h$44$^m$42.$^s$11 | +55°53′12.″7 | 0.0373 | 12.09[c] | (2) |
| IRAS 14348-1447 | 14$^h$37$^m$38.$^s$40 | −15°00′20.″0 | 0.0823 | 12.30[c] | (2) |
| IRAS 14394+5332 | 14$^h$41$^m$04.$^s$38 | +53°20′08.″7 | 0.1050 | 12.04[d] | (3) |
| IRAS 17179+5444 | 17$^h$18$^m$54.$^s$40 | +54°41′48.″5 | 0.1476 | 12.20[e] | (3) |
| IRAS F23529-2119 | 23$^h$55$^m$33.$^s$00 | −21°03′08.″7 | 0.4285 | 12.52[b] | (1) |
| IRAS 23389+0300 | 23$^h$41$^m$30.$^s$31 | +03°17′26.″4 | 0.1450 | 12.09[e] | (3) |

**Note.** Columns: (1) source name, (2) R.A., (3) decl., (4) spectroscopic redshift from NED (https://ned.ipac.caltech.edu/), (5) value of the absolute FIR luminosity with the reference in superscript, (6) reference for the source selection.
**References:** (1) Galvin et al. (2016), (2) Clemens et al. (2010), (3) Nandi et al. (2021).
[a] Spoon et al. (2009)
[b] Braun et al. (2011)
[c] Clemens et al. (2010)
[d] Kim & Sanders (1998)
[e] Veilleux et al. (1999)

We further note that Clemens et al. (2010) used a simple model, i.e., synchrotron+FFE+FFA, or a combination of these, to describe the radio SEDs of their sample. Moreover, Nandi et al. (2021) modeled the radio SEDs of these galaxies with the SYNAGE spectral fitting tool, which describes the evolution of a radio AGN continuously replenished by a constant flow of fresh relativistic particles with a PL energy distribution (Murgia et al. 1999). This tool works well for SED modeling of radio galaxies with large-scale structures as it assumes a cylindrical geometry of the sources and the equipartition of energy between the energetic particles and magnetic field. In addition, the flattening at low frequencies and steepening at high frequencies are interpreted as being due to SSA and particle energy losses. We reiterate that our sample is based on the availability of good data coverage at the UV–IR range, allowing us to perform simultaneous UV–IR–radio modeling using the CIGALE and radio-only SED modeling using dense radio SEDs and comparing the astrophysical properties inferred by these tools.

## 3. Multiwavelength Data Sets: Radio–FUV

### 3.1. Radio Flux Densities from the Literature

In order to construct dense radio SEDs for the selected targets, we further searched for flux measurements in the literature. In particular, we obtained measurements at the central frequencies from different sky surveys centered at 54 MHz from the LOFAR LBS Sky Survey (LoLSS) II (de Gasperin et al. 2023), 74 MHz from the VLA Low-Frequency Sky Survey (VLSSr; Cohen et al. 2007), 74–231 MHz from the GaLactic, and Extragalactic MWA Survey (GLEAM; Wayth et al. 2015), 150 MHz from the TIFR GMRT Sky Survey (TGSS) alternate data release (ADR) 1 (TGSS ADR1; Intema et al. 2017), 150 MHz from the LOw-Frequency ARray Two-meter Sky Survey (LoTSS) data release (DR)2 (Shimwell et al. 2022), 352 MHz from the Westerbork In the Southern Hemisphere (WISH; De Breuck et al. 2002), 365 MHz from the Texas survey (Douglas et al. 1996), 408 MHz from the Molonglo Reference Catalog (Large et al. 1981) 843 MHz from the Sydney University Molonglo Sky Survey (SUMSS; Mauch et al. 2003, 2013), 889 MHz from the Rapid ASKAP Continuum Survey (Hale et al. 2021), 1.4 GHz from the NRAO VLA Sky Survey (NVSS; Condon et al. 1998), 3.0 GHz from the VLA Sky Survey (VLASS; Lacy et al. 2020; Gordon et al. 2021), and 4.8 GHz from the Parkes-MIT-NRAO survey (Griffith & Wright 1993) within the positional uncertainties provided by the survey parameters.

Furthermore, since all of our galaxies have been targets for dedicated radio continuum monitoring, we included the measurements from either the GMRT (1.28 GHz; Nandi et al. 2021), or the Very Large Array (VLA; 235 MHz, 610 MHz, 15 GHz, and 22 GHz; Clemens et al. 2008, 2010) or ATCA (2.1, 5.5, and 9 GHz; Galvin et al. 2016) telescopes. We note that for the three galaxies from the Galvin et al. (2016) sample, IRAS F00183-7111, IRAS F0358-6432, and IRAS F23529-2119, the radio measurements at mid-frequency range were provided by ATCA, centered at 2.1, 5.5, and 9 GHz frequencies with a bandwidth of 2.04 GHz. For some targets, if enough signal-to-noise ratio (S/N >8) was available in a particular point-source model of each image, then the corresponding uv-data file was divided into two and so on to produce images in subbands and the subsequent flux densities are given (Table A2 of Galvin et al. 2016). For our radio SED modeling, we averaged the fluxes in a particular band, and those fluxes were used in the radio SED modeling.

Additionally, we used flux measurement provided from studies targeting individual sources such as IRAS F00183-7111 at 1.4 GHz (Drake et al. 2004), IRAS 12112+0305 and IRAS 10565+2448 at 8.4 GHz (Vardoulaki et al. 2015), IRAS F00183-7111 at 4.8 and 8.6 GHz (McConnell et al. 2012), IRAS 23389+0300 at 178 MHz (Gower et al. 1967),

IRAS 08572+3915 at 32 GHz (Barcos-Muñoz et al. 2017), UGC 5101 at 325 MHz (Rengelink et al. 1997) and 32.5 GHz (Leroy et al. 2011), and UGC 08058 at 30.0 GHz (Lowe et al. 2007) and 32.5 GHz (Leroy et al. 2011).

The GLEAM survey provides data in the frequency bands of 72–103 MHz, 103–134 MHz, 139–170 MHz, 170–200 MHz, and 200–231 MHz, where each band is divided into 7.68 MHz subbands for imaging purposes (Hurley-Walker et al. 2017). However, the flux measurements from each subband are not independent of each other; therefore, we averaged the fluxes within each subband for spectral modeling, providing a frequency resolution of 30.72 MHz.

### 3.2. Archival VLA Data

For two galaxies in our sample, IRAS 14394+5332 and IRAS 23389+0303, we found additional radio archival VLA data at 8.4 GHz. The two targets were observed on 2004 October 19, under project ID AN 122, in the A-array configuration for a total duration of 22 and 20.25 minutes, respectively. We calibrated the interferometric data using the NRAO AIPS package following the standard procedure. The flux density scale in Baars et al. (1977) was used to obtain the flux densities of the primary (flux) calibrator (1331+305), the secondary (phase) calibrator (1419+543 and 2320+052 for IRAS 14394+5332 and IRAS 23389+0303, respectively), and the target source. Antennas and baselines affected by strong radio frequency interference and nonworking antennas were removed after visual inspection. Images were produced using the task IMAGR on the data after both frequency channels were calibrated separately and averaged. Usually, three to five rounds of phase-only self-calibration were performed iteratively by choosing point sources in the field such that the flux density was $>3\sigma$ with one synthesized beam. The final images were made with full UV coverage and a robust weighting of 0 (Briggs 1995). The final images were corrected for the reduction in sensitivity, due to the shape of the antenna primary beam, using the task PBCOR.[5] Integrated flux densities (and uncertainty) were obtained using the task TVSTAT in AIPS. Figure A1 in Appendix A provides the radio optical overlays and preliminary information on the data sets.

Table A1 summarizes the integrated flux densities and their uncertainties, upscaled to account for variations in uncalibrated system temperature. In particular, we added in quadrature the flux density errors (see Equation (1) of Żywucka et al. 2014): an additional 6% for LoLLS (de Gasperin et al. 2023), 5% for VLSSr (Cohen et al. 2007), 10% for 150 MHz TGSS ADR1 (Intema et al. 2017), 10% for 150 MHz LoTSS (Shimwell et al. 2022), 5% for 327 MHz WENSS (Hardcastle et al. 2016), 13% for 352 MHz WISH (De Breuck et al. 2002), 365 MHz Texas survey (Douglas et al. 1996), 5% for 1.28 GHz data from Nandi et al. (2021), 10% for ASKAP (Hale et al. 2021), 3% for 4.8, 8.4, 14.9, and 32.5 GHz observations from VLA (Perley & Butler 2017), 3% for NVSS (Condon et al. 1998), 10% for VLASS (Lacy et al. 2020), 10% for GLEAM (Mhlahlo & Jamrozy 2021), 5% for SUMSS (Mauch et al. 2003; Galvin et al. 2018), 5% for ATCA (Galvin et al. 2018), and 0.2% for 32.0 GHz (Lowe et al. 2007) measurements.

We note that the synthesized beam size for the majority of our sources ranges from a few arcsecs to a few tens of arcsecs (except for the VLA measurements where the synthesized beam sizes are of the order of subarcsec at 8.4 GHz for the data analyzed by us; Figure A1), the linear scale corresponds to ∼3–30 kpc for the z-range of 0.0373–0.428 for our sources, assuming a typical resolution of 5″. Therefore, it is reasonable to assume that the integrated fluxes used in our SED modeling are from the extended regions of our sources.

### 3.3. UV, Optical, and IR Flux Densities

We gathered photometric measurements from several instruments from ground- and space-based facilities for SED modeling using CIGALE. Specifically, we crossmatched the positions of our galaxies with those provided in public databases such as the NASA/IPAC Extragalactic Database (NED 2019), SIMBAD (Wenger et al. 2000), VizieR (Ochsenbein 1996), and NASA/IPAC Infrared Science Archive (IRSA)[6] using a matching radius of 5″–15″.

About 20–30 bands of UV–IR broadband photometry are available for each of these sources. They include measurements from the Galaxy Evolution Explorer (GALEX), Optical/UV monitor of XMM-Newton telescope (XMM-OM), Swift ultraviolet/optical telescope, SkyMapper Southern Sky Survey (SMSS), Sloan Digital Sky Survey (SDSS) DR 16, the Two-Micron All-Sky Survey (2MASS), the Wide-field Infrared Survey Explorer (WISE), Spitzer space telescope, AKARI, and Herschel Space Observatory. Regarding the availability of multiple flux measurements at a given wavelength, we chose the one that contained the entire galaxy. For interacting or merging-type galaxies, the fluxes used in modeling include emission from the companion, depending on the wavelength used and the resolution of the survey. Table B1 gives the integrated fluxes and the integration area per band for each source used in the SED modeling.

We note that an aperture mismatch between different instruments can result in inconsistent flux measurement and influence the final SED. The resulting SED fitting can be affected by the decrement in the flux when the beam size is smaller than the galaxy itself or by a larger aperture, which can result in measuring a partial flux from a nearby galaxy. However, for our sample of galaxies, we tried to use an aperture size that corresponded well to the size of the galaxy. It is not possible to match it perfectly without remeasurement of the fluxes, and we have not done so in our analysis (for improvement in SED quality, see Ramos Padilla et al. 2020). Moreover, the parameters derived from the SED fitting tools are in agreement when apertures are matched with those obtained without matching the apertures (Ramos Padilla et al. 2020). During SED modeling, CIGALE adds an additional error of 10% of the flux in quadrature together with the existing errors. This additional 10% uncertainty counts for all possible uncertainties, for example, aperture, that is not included by default in the photometric error bars. Considering the wavelength coverage of the analyzed ULIRGs, we are confident that even with small aperture problems for selected bands, the final fitted SED mimics the overall estimated physical parameters very well. Data ranging from the near UV to the optical, IR, and radio ensure that the obtained models describe the galaxy very well, without any possible misclassification between passive/star-forming, non-AGN/AGN, etc. The Bayesian analysis, based on millions of templates generated by the SED fitting method, gives us not only the

---

[5] http://www.aips.nrao.edu/cgi-bin/ZXHLP2.PL?PBCOR

[6] https://irsa.ipac.caltech.edu/

Table 2
Radio Continuum Models Considered for Spectral Fitting

| Model Name | Radio Spectrum | Free Parameters |
|---|---|---|
| Single Emission Component PL | | |
| | $S_\nu = A\left(\frac{\nu}{\nu_0}\right)^\alpha$ | $A, \alpha$ |
| Synchrotron and FFE (SFG NC) | $S_\nu = A\left(\frac{\nu}{\nu_0}\right)^\alpha + B\left(\frac{\nu}{\nu_0}\right)^{-0.1}$ | $A, B, \alpha$ |
| Synchrotron and FFE with FFA (C) | $S_\nu = (1 - e^{-\tau})\left[B + A\left(\frac{\nu}{\nu_{t,1}}\right)^{0.1+\alpha}\right]\left(\frac{\nu}{\nu_{t,1}}\right)^2$ | $A, B, \nu_{t,1}, \alpha$ |
| Multiple Emission Components SFG NC2 | | |
| | $S_\nu = A\left(\frac{\nu}{\nu_0}\right)^{\alpha_1} + B\left(\frac{\nu}{\nu_0}\right)^{-0.1} +$ | $A, B, \alpha_1,$ |
| | $C\left(\frac{\nu}{\nu_0}\right)^{\alpha_2} + D\left(\frac{\nu}{\nu_0}\right)^{-0.1}$ | $C, D, \alpha_2$ |
| C2 1SA | $S_\nu = (1 - e^{-\tau_1})\left[B + A\left(\frac{\nu}{\nu_{t,1}}\right)^{0.1+\alpha}\right]\left(\frac{\nu}{\nu_{t,1}}\right)^2 +$ | $A, B, C, D,$ |
| | $(1 - e^{-\tau_2})\left[D + C\left(\frac{\nu}{\nu_{t,2}}\right)^{0.1+\alpha}\right]\left(\frac{\nu}{\nu_{t,2}}\right)^2$ | $\alpha, \nu_{t,1}, \nu_{t,2}$ |
| C2 1SAN | $S_\nu = \left(\frac{\nu}{\nu_0}\right)^{-2.1}\left[B + A\left(\frac{\nu}{\nu_0}\right)^{0.1+\alpha}\right]\left(\frac{\nu}{\nu_0}\right)^2 +$ | $A, B, C, D,$ |
| | $(1 - e^{-\tau_2})\left[D + C\left(\frac{\nu}{\nu_{t,2}}\right)^{0.1+\alpha}\right]\left(\frac{\nu}{\nu_{t,2}}\right)^2$ | $\alpha, \nu_{t,2}$ |
| C2 1SAN2 | $S_\nu = \left(\frac{\nu}{\nu_0}\right)^{-2.1}\left[B + A\left(\frac{\nu}{\nu_0}\right)^{0.1+\alpha_1}\right]\left(\frac{\nu}{\nu_0}\right)^2 +$ | $A, B, C, D,$ |
| | $(1 - e^{-\tau_2})\left[D + C\left(\frac{\nu}{\nu_{t,2}}\right)^{0.1+\alpha_2}\right]\left(\frac{\nu}{\nu_{t,2}}\right)^2$ | $\alpha_1, \alpha_2, \nu_{t,2}$ |
| C2 | $S_\nu = (1 - e^{-\tau_1})\left[B + A\left(\frac{\nu}{\nu_{t,1}}\right)^{0.1+\alpha_1}\right]\left(\frac{\nu}{\nu_{t,1}}\right)^2 +$ | $A, B, C, D,$ |
| | $(1 - e^{-\tau_2})\left[D + C\left(\frac{\nu}{\nu_{t,2}}\right)^{0.1+\alpha_2}\right]\left(\frac{\nu}{\nu_{t,2}}\right)^2$ | $\alpha_1, \alpha_2, \nu_{t,1}, \nu_{t,2}$ |

**Note.** $A, C$ and $B, D$ are the normalization components for the synchrotron and FFE, respectively. $\alpha, \alpha_1,$ and $\alpha_2$ are the PL indices for the synchrotron emission. $\nu_{t,1}$ and $\nu_{t,2}$ are the turnover frequencies from the FFA, while $\tau$ gives the optical depth for the thermal gas. $\nu_0$ gives the reference frequency, set to 1.4 GHz.

uncertainty of the estimated parameters, but also their probability distribution functions (PDFs; Boquien et al. 2019). In our CIGALE analysis, we checked all PDFs for estimated physical properties, such as stellar masses, SFR, dust mass etc., which showed a single peak profile indicative of their robust estimations (Boquien et al. 2019). Indeed, the aperture problem with very limited photometric coverage of the galaxy spectra can affect the final result (see Małek et al. 2018). However, in the case of selected 14 ULIRGs with very dense coverage of the spectra, the aperture inhomogeneity can affect the uncertainty but not the main physical parameters analyzed in this paper.

## 4. SED Modeling

### 4.1. Radio SED Modeling

In accordance with our previous study involving SED modeling of LIRGs, we also modeled the integrated radio SEDs for the sample of ULIRGs with physically motivated scenarios. The methodology is described in detail in Dey et al. (2022) and is presented in Table 2. Briefly, these models include:

1. A single volume of emission plasma consisting of thermal free–free plasma intermixed with synchrotron emission producing relativistic electrons. Synchrotron emission may be attenuated by FFA, causing low-frequency turnover, which depends on the density and distribution of ionized electrons concerning the synchrotron emission. This volume represents a single star-forming region with one composition (optical depth $\tau$) and is characterized by a single electron population ($\alpha$). We distinguish the following three scenarios within this framework:
   (a) A synchrotron component described by a single PL.
   (b) A synchrotron and FFE components without low-frequency curvature (SFG NC).
   (c) A synchrotron and FFE components with FFA including a single low-frequency turnover characterized by optical depth, $\tau$, defined as $(\nu/\nu_{t,1})^{-2.1}$, where $\nu_{t,1}$ represents the turnover frequency due to FFA (C).

2. Multiple (two) volumes of emission plasma, including the synchrotron and FFA emission components and absorption features representing multiple star-forming regions with different orientations or compositions (optical depths, $\tau_1$ and $\tau_2$), characterized by a single electron population ($\alpha$) or different electron populations ($\alpha_1$ and $\alpha_2$). We distinguish the following five scenarios within this framework:
   (a) Synchrotron and FFE components without a low-frequency curvature (SFG NC2).
   (b) Synchrotron and FFE components with a single $\alpha$ having different optical depths, $\tau_1$ and $\tau_2$, respectively (C2 1SA).
   (c) Synchrotron and FFE components with a single $\alpha$ and no turnover at $\tau_1$ and a turnover at $\tau_2$ (C2 1SAN).
   (d) Synchrotron and FFE components with no turnover at $\tau_1$ and two different electron populations, $\alpha_1$ and $\alpha_2$ (C2 1SAN2).

Table 3
An Overview of the Natural Log of the Bayes Odds Ratio from the UltraNest Fitting of Each Model to Each Source

| Name | PL | SFG NC | C | SFG NC2 | C2 1SA | C2 1SAN | C2 1SAN2 | C2 |
|---|---|---|---|---|---|---|---|---|
| IRAS F00183-7111 | −143.04 | −143.91 | **0** | −152.04 | *−0.24* | −10.10 | −6.88 | −1.52 |
| IRAS F03538-6432 | **0** | *−6.72* | −17.65 | −13.53 | −10.21 | −13.53 | −14.47 | −10.91 |
| IRAS 08572+3915 | −102.64 | *−0.35* | **0** | −2.46 | −10.00 | −14.55 | −11.72 | −10.69 |
| UGC 5101 | −43.89 | −51.02 | −15.47 | −60.13 | −2.79 | −1.95 | **0** | −2.074 |
| IRAS 10565+2448 | −41.32 | −48.38 | **0** | −57.07 | −7.34 | −11.09 | −10.24 | −7.49 |
| IRAS 12112+0305 | −71.61 | −52.28 | **0** | −58.88 | −8.81 | −14.01 | −11.49 | −9.99 |
| UGC 08058 | *−636.75* | −87.51 | −839.90 | −351.81 | −5.70 | −3.24 | **0** | −12.84 |
| IRAS 13305-1739 | **0** | *−7.60* | −12.19 | −24.25 | −14.174 | −9.67 | −6.62 | −14.33 |
| UGC 8696 | −54.78 | −62.43 | −19.32 | −72.51 | −6.49 | −29.48 | −29.63 | **0** |
| IRAS 14348-1447 | **0** | *−7.29* | −20.77 | −14.04 | *−1.65* | −3.24 | *−1.83* | *−1.78* |
| IRAS 14394+5332 | −35.73 | −43.64 | −38.29 | −54.24 | *−1.75* | −65.61 | −2.79 | **0** |
| IRAS 17179+5444 | −419.38 | −255.02 | **0** | −269.93 | −5.85 | −23.79 | −15.42 | −5.99 |
| IRAS F23529-2119 | **0** | *−6.58* | −7.38 | −14.40 | *−1.93* | −4.72 | −7.36 | *−1.67* |
| IRAS 23389+0300 | −1143.71 | −49.87 | **0** | −56.48 | −6.40 | −11.60 | −9.09 | −7.58 |

**Note.** We present the most preferred model with the highest evidence value corresponding to loge(1) = 0 (shown in boldface). Models that are indistinguishable from the most preferred model correspond to loge(3) = 1.1 (shown in italics). Less preferred models have larger negative values.

(e) Synchrotron and FFE components with two $\alpha$s ($\alpha_1$, $\alpha_2$) and different optical depths, $\tau_1$, and $\tau_2$ (C2).

We used a Bayesian inference Python package called UltraNest (Buchner 2021) to derive the posterior probability distributions for the best-fit model parameters and the Bayesian evidence value for model comparison. In our analysis, we considered the uniform prior distribution of model parameters. We constrain the priors for the normalization parameters $A$, $B$, $C$, $D$ in positive ranges, the spectral index parameters ($\alpha$, $\alpha_1$ and $\alpha_2$) in the range of −0.2 and −2.0, and the turnover frequencies ($\nu_{t,1}$ and $\nu_{t,2}$) are between 1 MHz and 50 GHz.

The best model is selected based on the Bayes odds ratio logarithm and is given in Table 3, while the corresponding best-fit radio SEDs are given in Figure 1 for our sample. Among all the examined models, the one with the Bayes odds ratio value equal to 0 (corresponding to natural loge(1) = 0) is considered the best model given the data. However, the models are indistinguishable when the ratio is less than loge(3) = 1.1. Therefore, if two or more models turn out to be indistinguishable from each other, we select the best-fit model as the one with the least number of free parameters. Table 4 provides the 50th percentile of the posterior distribution of best-fit model parameters and their equivalent 1σ uncertainties corresponding to the 16th and 84th percentiles of the posterior distribution of the parameter. To evaluate the degeneracy arising from the multiple free parameters in the best-fit model, we present corner plots for each of our galaxies in Appendix A (Figure A2). The complete set of figures (comprising 14 corner plots) is available in the online journal. The corner plot presented in Figure A2 represents the one- and two-dimensional projections of the posterior probability distribution of parameters. Table 5 provides total, thermal, and nonthermal radio fluxes from SED modeling, along with the thermal fraction (TF) at 1.4 GHz (TF$_{1.4\,\text{GHz}}$), which was computed from the decomposed thermal and nonthermal flux estimated using the radio SED modeling.

### 4.2. UV–Radio SED Modeling with CIGALE

For simultaneous UV–radio SED modeling of our galaxies, we used the recently updated version of the X-CIGALE software (Yang et al. 2020), CIGALE 2022.1 (Yang et al. 2022). This code works on the principle of energy balance, i.e., attenuation of energy in UV to optical being self-consistently reemitted in IR. Its parallelized and modular user-friendly nature makes it widely used in the literature to estimate the astrophysical properties of targets. Below we provide a brief description of the models used to model the galaxy emission, while a comprehensive discussion about the models is provided in Dey et al. (2022). We used the Bruzual & Charlot (2003) single stellar population template to model the stellar emission considering a Salpeter (1955) IMF. Furthermore, we modeled the star formation history with a delayed SFR with an additional burst profile (in accordance with Małek et al. 2018). The attenuation of stellar emission by dust is modeled by Calzetti et al. (2000), and dust emission is modeled with The Heterogeneous Evolution Model for Interstellar Solids (THEMIS; Jones et al. 2017). The AGN emission is modeled with the SKIRTOR module, which takes into account the clumpiness of the torus (Stalevski et al. 2012, 2016). Further, CIGALE modeling is done in a rest frame where the different modules allow for the building of a full FUV–radio rest-frame spectrum. The subsequent SED is built in the observer's frame by applying the redshifting module, which redshifts the spectrum and dims it, multiplying the wavelengths by $1 + z$ and dividing the spectrum by $1 + z$. For the K-correction in CIGALE, the spectral index is assumed to be 0, which is reasonable as the difference between mean magnitudes in the observed frame WFIR band and rest-frame SDSS $r$ band is found to be close to 0 in the redshift $z < 0.5$ for the average spiral and elliptical galaxy types (Wolf et al. 2003). The physical properties, such as SFR, attenuation, stellar, dust mass, etc., are computed using the full spectrum before redshifting it (Boquien et al. 2019).

The advantage of CIGALE 2022.1 as compared to its predecessors is that here the radio module includes both the thermal (FFE from the nebular emission) and the nonthermal emission (synchrotron emission from the star-forming and AGN component) of the galaxy. The addition of a new AGN component to the radio module of CIGALE 2022.1 allowed us to estimate the radio-loudness, $R_{\text{AGN}}$, defined as a ratio between AGN luminosities measured at 5 GHz and 2500 Å, and then the

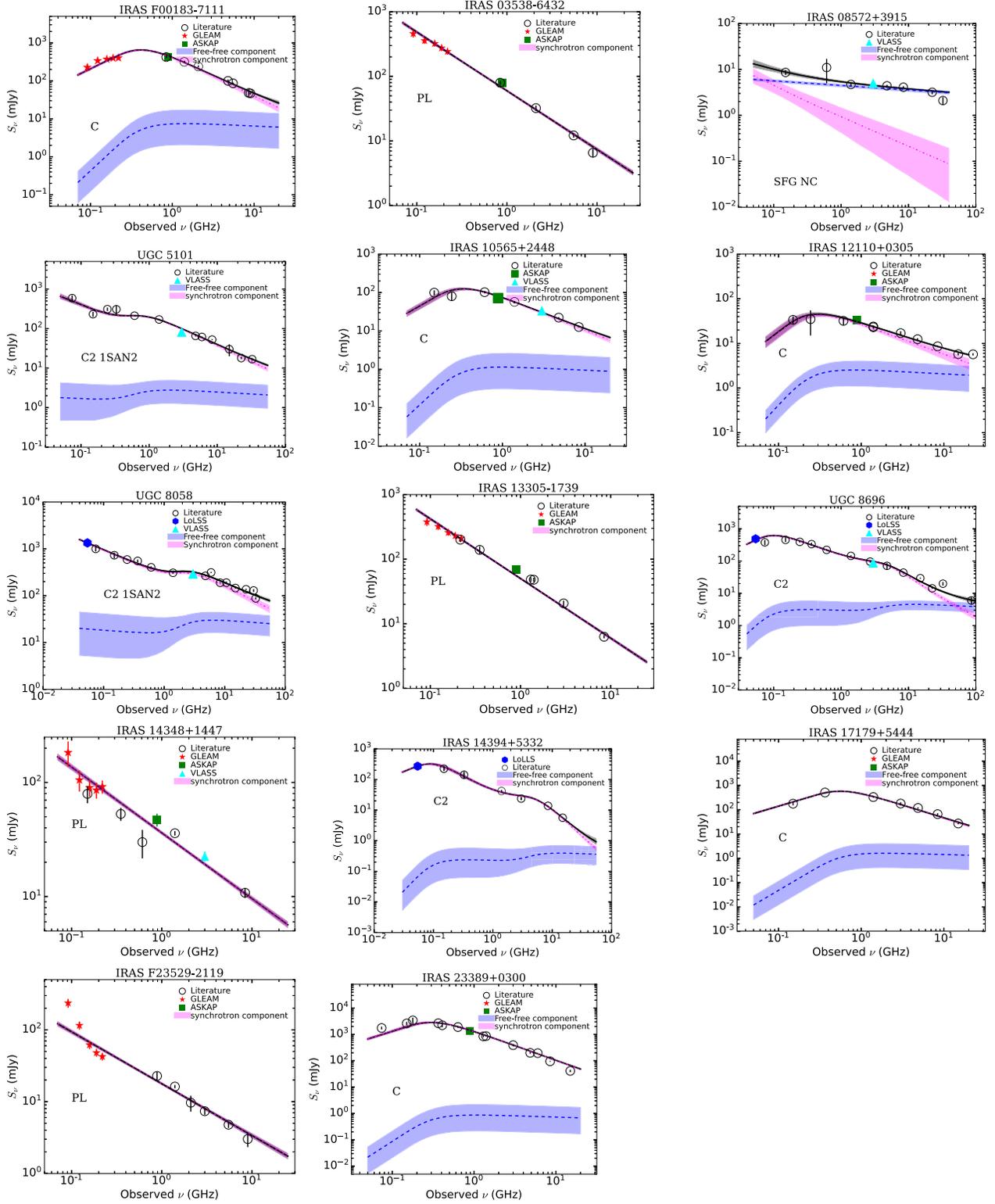

**Figure 1.** Radio SEDs for our sample of 14 ULIRGs. Galaxy name and the best-fit radio SED model name are given at the top and bottom left corner, respectively, of each panel. The solid black line represents the best-fit model, while the gray shaded region represents the $1\sigma$ uncertainties sampled by the UltraNest package. The dotted–dashed magenta and dotted blue lines show the decomposed synchrotron and free–free components. The pink and blue shaded regions represent the $1\sigma$ uncertainties in the synchrotron and free–free components, respectively. For galaxies fitted with two-component models, the free–free and synchrotron components correspond to the sum of individual free–free and synchrotron components. The flux densities obtained from the GLEAM, ASKAP, and VLASS surveys are shown with red stars, green squares, and cyan triangles, respectively, while open black circles mark the fluxes taken from other studies. (see Section 3.1, Table A1).

slope of the PL AGN radiation (assumed isotropic; Yang et al. 2022). To maintain consistency with our previous study (Dey et al. 2022), we used the same modules. CIGALE 2022.1 takes into account the radio luminosities by star formation and AGN components using the $q_{IR}$ and radio-loudness, $R_{AGN}$, parameter scaling relations. At 1.4 GHz, the radio luminosity from star

**Table 4**
The Most Preferred Models Were Selected Based on the Bayes Odds Ratio and the Constrained Value of the Free Parameters

| Name | Model | $A$ (mJy) | $B$ (mJy) | $\alpha(=\alpha_1)$[a] | $\nu_{t,1}$ (GHz) | $C$ (mJy) | $D$ (mJy) | $\alpha_2$ | $\nu_{t,2}$ (GHz) |
|---|---|---|---|---|---|---|---|---|---|
| (1) | (2) | (3) | (4) | (5) | (6) | (7) | (c) | (9) | (10) |
| IRAS F00183-7111 | C | $1004.84^{+35.52}_{-38.54}$ | $9.60^{+12.30}_{-7.00}$ | $-1.05^{+0.06}_{-0.09}$ | $0.45^{+0.04}_{-0.03}$ | ⋯ | ⋯ | ⋯ | ⋯ |
| IRAS F03538-6432 | PL | $44.08^{+1.55}_{-1.46}$ | ⋯ | $-0.91^{+0.02}_{-0.02}$ | ⋯ | ⋯ | ⋯ | ⋯ | ⋯ |
| IRAS 08572+3915 | SFG NC | $0.80^{+0.44}_{-0.43}$ | $4.29^{+0.29}_{-0.32}$ | $-0.62^{+0.10}_{-0.25}$ | ⋯ | ⋯ | ⋯ | ⋯ | ⋯ |
| UGC 5101 | C2 1SAN2 | $50.67^{+13.40}_{-15.95}$ | $1.38^{+1.96}_{-1.01}$ | $-0.76^{+0.09}_{-0.15}$ | ⋯ | $207.63^{+34.13}_{-27.67}$ | $1.51^{+1.99}_{-1.10}$ | $-0.88^{+0.07}_{-0.11}$ | $0.92^{+0.13}_{-0.11}$ |
| IRAS 10565+2448 | C | $196.79^{+13.89}_{-13.11}$ | $1.25^{+1.87}_{-0.92}$ | $-0.86^{+0.04}_{-0.05}$ | $0.33^{+0.04}_{-0.04}$ | ⋯ | ⋯ | ⋯ | ⋯ |
| IRAS 12112+0305 | C | $67.93^{+6.72}_{-6.23}$ | $2.93^{+1.74}_{-1.75}$ | $-0.67^{+0.07}_{-0.10}$ | $0.26^{+0.05}_{-0.04}$ | ⋯ | ⋯ | ⋯ | ⋯ |
| UGC 08058 | C2 1SAN2 | $212.06^{+15.78}_{-17.29}$ | $13.50^{+17.02}_{-9.94}$ | $-0.56^{+0.02}_{-0.02}$ | ⋯ | $241.74^{+19.70}_{-19.26}$ | $15.74^{+17.76}_{-11.42}$ | $-0.75^{+0.08}_{-0.10}$ | $2.88^{+0.09}_{-0.16}$ |
| IRAS 13305-1739 | PL | $36.78^{+0.95}_{-1.03}$ | ⋯ | $-0.92^{+0.01}_{-0.02}$ | ⋯ | ⋯ | ⋯ | ⋯ | ⋯ |
| UGC 8696 | C2 | $932.94^{+36.39}_{-34.66}$ | $3.75^{+3.37}_{-2.56}$ | $-0.93^{+0.04}_{-0.05}$ | $0.10^{+0.003}_{-0.002}$ | $61.25^{+9.95}_{-9.48}$ | $2.05^{+2.09}_{-1.47}$ | $-1.92^{+0.15}_{-0.06}$ | $5.87^{+0.69}_{-0.69}$ |
| IRAS 14348-1447 | PL | $29.81^{+0.92}_{-0.88}$ | ⋯ | $-0.58^{+0.02}_{-0.02}$ | ⋯ | ⋯ | ⋯ | ⋯ | ⋯ |
| IRAS 14394+5332 | C2 | $452.12^{+19.82}_{-20.61}$ | $0.30^{+0.41}_{-0.22}$ | $-1.24^{+0.09}_{-0.12}$ | $0.10^{+0.01}_{-0.01}$ | $25.11^{+3.22}_{-3.12}$ | $0.18^{+0.25}_{-0.13}$ | $-1.97^{+0.05}_{-0.03}$ | $6.30^{+0.42}_{-0.43}$ |
| IRAS 17179+5444 | C | $900.09^{+23.66}_{-25.39}$ | $2.05^{+2.97}_{-1.55}$ | $-1.09^{+0.04}_{-0.04}$ | $0.63^{+0.04}_{-0.03}$ | ⋯ | ⋯ | ⋯ | ⋯ |
| IRAS F23529-2119 | PL | $13.95^{+0.47}_{-0.50}$ | ⋯ | $-0.72^{+0.02}_{-0.02}$ | ⋯ | ⋯ | ⋯ | ⋯ | ⋯ |
| IRAS 23389+0300 | C | $4513.09^{+212.85}_{-181.97}$ | $0.92^{+1.38}_{-0.69}$ | $-1.10^{0.01}_{0.01}$ | $0.32^{+0.02}_{-0.02}$ | ⋯ | ⋯ | ⋯ | ⋯ |

**Note.** The nominal value is taken as the 50th percentile of the posterior distribution of the samples, and the 1σ uncertainties are provided by the 16th and 84th percentiles. Parameters not included in the models are indicated as "⋯." Columns: (1) galaxy name, (2) best-fit model, (3) and (7) nonthermal normalization components, (4) and (8) thermal normalization components, (5) and (9) synchrotron spectral indices, (6) and (10) turnover frequencies.
[a] $\alpha = \alpha_1$ for the two-component models.

**Table 5**
Total, Thermal, and Nonthermal Fluxes Are Estimated at 1.4 GHz from the Radio-only SED Fitting, and the Corresponding TF Is Estimated at 1.4 GHz

| Name | $S_{1.4}^{total}$ (mJy) | $S_{1.4}^{th}$ (mJy) | $S_{1.4}^{nth}$ (mJy) | $TF_{1.4}$ |
|---|---|---|---|---|
| (1) | (2) | (3) | (4) | (5) |
| IRAS F00183-7111 | $302.88^{+10.87}_{-10.71}$ | $8.59^{+6.31}_{-9.84}$ | $293.40^{+15.65}_{-13.31}$ | $0.03^{+0.02}_{-0.03}$ |
| IRAS F03538-6432 | $44.12^{+1.42}_{-1.42}$ | ⋯ | $44.12^{+1.42}_{-1.42}$ | ⋯ |
| IRAS 08572+3915 | $5.07^{+0.23}_{-0.23}$ | $4.32^{+0.31}_{-0.30}$ | $0.75^{+0.45}_{-0.46}$ | $0.86^{+0.10}_{-0.09}$ |
| UGC 5101 | $171.31^{+5.38}_{-5.26}$ | $3.17^{+1.74}_{-2.28}$ | $167.82^{+4.94}_{-5.08}$ | $0.02^{+0.01}_{-0.01}$ |
| IRAS 10565+2448 | $57.05^{+1.76}_{-1.76}$ | $1.10^{+0.81}_{-1.46}$ | $55.74^{+2.27}_{-2.05}$ | $0.02^{+0.01}_{-0.02}$ |
| IRAS 12112+0305 | $23.90^{+0.90}_{-0.79}$ | $2.59^{+1.48}_{-1.36}$ | $21.28^{+1.52}_{-1.62}$ | $0.11^{+0.06}_{-0.05}$ |
| UGC 8058 | $281.27^{+7.80}_{-8.16}$ | $13.27^{+8.23}_{-13.48}$ | $305.27^{+14.41}_{-11.96}$ | $0.05^{+0.03}_{-0.05}$ |
| IRAS 13305-1739 | $36.77^{+1.05}_{-0.98}$ | ⋯ | $36.77^{+1.05}_{-0.98}$ | ⋯ |
| UGC 8696 | $131.82^{+4.27}_{-4.36}$ | $0.98^{+0.70}_{-1.62}$ | $130.47^{+4.55}_{-4.55}$ | $0.007^{+0.004}_{-0.008}$ |
| IRAS 14348-1447 | $29.76^{+0.91}_{-0.87}$ | ⋯ | $29.76^{+0.91}_{-0.87}$ | ⋯ |
| IRAS 14394+5332 | $39.77^{+2.38}_{-2.32}$ | $0.27^{+0.19}_{-0.35}$ | $39.42^{+2.43}_{-2.35}$ | $0.007^{+0.005}_{-0.009}$ |
| IRAS 17179+5444 | $347.11^{+13.07}_{-13.08}$ | $1.65^{+1.22}_{-2.42}$ | $344.98^{+12.94}_{-12.96}$ | $0.005^{+0.003}_{-0.007}$ |
| IRAS F23529-2119 | $13.94^{+0.45}_{-0.46}$ | ⋯ | $13.94^{+0.45}_{-0.46}$ | ⋯ |
| IRAS 23389+0300 | $881.21^{+9.51}_{-10.49}$ | $0.78^{+0.58}_{-1.17}$ | $880.12^{+9.70}_{-10.50}$ | $0.0008^{+0.0006}_{-0.0013}$ |

**Note.** The nominal value is taken as the 50th percentile of the posterior distribution of the samples, and the 1σ uncertainties are provided by the 16th and 84th percentiles. Columns: (1) galaxy name, (2–4) total, thermal, and nonthermal fluxes at 1.4 GHz, (5) TF at 1.4 GHz.

formation is dominated by nonthermal synchrotron emission, and therefore, calculated using the $q_{IR}$ parameter. The $q_{IR}$ sets the normalization at this frequency, and a single PL form describes the synchrotron emission from star formation. The scatter and variations in the $q_{IR}$ parameter itself will contribute to the uncertain modeling of the radio component by the CIGALE. The nonthermal synchrotron emission from the AGN, described by the $R_{AGN}$ parameters, also assumes the single PL form. However, radio SEDs are rarely characterized by single-PL forms (see Section 1). Therefore, while including radio measurements in CIGALE modeling improves the accuracy of derived parameters, it does not fully compensate for the need for radio-only SED modeling using densely sampled radio SEDs, which is crucial for understanding galaxies' emission and absorption processes. Table 6 provides the parameter grid of the models used by us and Table 7 gives the best-fit parameters of the models. We stress here that the large number of photometric data points with uniform sampling across the wavelength range allows us to use a dense grid of parameters without the risk of overfitting model parameters.

**Table 6**
List of Input Parameters Used for CIGALE Modeling

| Parameters | | Values |
|---|---|---|
| Delayed Star Formation History + Additional Burst (Ciesla et al. 2015) | | |
| e-folding time of the main stellar population model [Myr] | $\tau_{main}$ | 300–15,000 by a bin of 300 |
| e-folding time of the late starburst population model [Myr] | $\tau_{burst}$ | 50–1000 by a bin of 50 |
| Mass fraction of the late burst population | $f_{burst}$ | 0.05, 0.1, 0.3, 0.6, 0.9 |
| Age of the main stellar population in the galaxy [Myr] | age | 1000, 2000, 3000, 4500, 5000, 6500, 10,000, 12,000 |
| Age of the late burst [Myr] | age$_{burst}$ | 10.0, 40.0, 80.0, 110, 150, 170 |
| Stellar Synthesis Population (Bruzual & Charlot 2003) | | |
| Initial mass function | IMF | (Salpeter 1955) |
| Metallicity | Z | 0.02 |
| Separation age | | 1 Myr |
| Dust Attenuation Laws (Calzetti et al. 2000) | | |
| Color excess of young stars | $E(B-V)$ | 0.1–2 by a bin of 0.2 |
| Reduction factor [iii] | $f_{att}$ | 0.3, 0.44, 0.6, 0.7 |
| Dust Grain Model: THEMIS (Jones et al. 2017) | | |
| Fraction of small hydrocarbon solids | $q_{hac}$ | 0.02, 0.06, 0.1, 0.17, 0.24 |
| Minimum radiation field | $U_{min}$ | 1, 5, 10, 15, 20, 30 |
| PL index of the radiation | $\alpha$ | 2 |
| Fraction illuminated from $U_{min}$ to $U_{max}$ | $\gamma$ | 0.02, 0.06, 0.1, 0.15, 0.2 |
| Active Nucleus Model: Skirtor (Stalevski et al. 2012, 2016) | | |
| Optical depth at 9.7 $\mu$m | $\tau_{9.7}$ | 3.0, 7.0 |
| Torus density radial parameter | pl | 1.0 |
| Torus density angular parameter | q | 1.0 |
| Angle between the equatorial plane and edge of the torus [deg] | oa | 40.0 |
| Ratio of outer to inner radius | R | 20.0 |
| Fraction of total dust mass inside clumps [%] | Mcl | 97.0 |
| Inclination (viewing angle) [deg] | i | 30 (type 1), 70 (type 2) |
| AGN fraction | | 0.0–0.4 by a bin of 0.05 |
| Extinction law of polar dust | | SMC |
| $E(B-V)$ of polar dust | | 0.01–0.7 by a bin of 0.5 |
| Temperature of the polar dust | K | 100 |
| Emissivity index of the polar dust | | 1.6 |
| Radio | | |
| Star formation FIR/radio parameter[a] | $q_{IR}$ | 2.3–2.9 by a bin of 0.1 |
| Star formation PL slope (flux $\propto$ frequency$^{\alpha_{synch}}$) | $\alpha_{SF}$ | −2.0 to −0.2 by a bin of 0.1 |
| Radio-loudness parameter[b] | $R_{AGN}$ | 0.01, 0.05, 0.1, 0.5... 1000, 5000 |
| AGN PL slope | $\alpha_{AGN}$ | −2.0 to −0.2 by a bin of 0.1 |

**Notes.**
[a] Computed as $\log_{10} L_{IR(8-1000\mu m)}$ -$\log_{10} L_{1.4\,GHz}$, where $L_{1.4\,GHz}$ is the radio luminosity at 1.4 GHz.
[b] Defined as the ratio between AGN luminosities measured at 5 GHz and 2500 Å,

## 5. Results

Our main results using detailed radio-only and simultaneous FUV–IR–radio SED modeling using the UltraNest and CIGALE modeling tools, respectively, are given below. The specific notes on individual sources are given in Appendix C. Figures 1 and 2 present the radio-only and FUV–radio SED fits, respectively. Tables 4 and 7 give the best-fit model parameters and their uncertainties, respectively, for our galaxies. Our main results are the following:

1. Radio-only SED modeling of 14 ULIRGs has been carried out in previous studies (Section 2); however, in this study, we obtained the best-fit radio SED model from the pool of physically motivated models using the state-of-the-art fitting tool (Figure 1). Most galaxies are fitted with complex models instead of single PL forms. The estimated radio spectral index ranges between −0.58 and −1.94 (Table 4).

2. Spectral index values between ∼−0.6 and ∼−0.9 can be considered within the uncertainty, in agreement with the canonical synchrotron emission spectral index ($\alpha \approx$ −0.75 for the frequently assumed value of the particle energy index = −2.5; Lacki et al. 2010). Values between ∼−1.0 and ∼−1.5 can be considered according to the cooled synchrotron emission spectral index ($\alpha$ = −1.25). For two galaxies (UGC 8696 and IRAS 14394+5332), the spectra are best fitted with a two-component model, where the derived spectral index of the second component is much steeper, ∼−1.9. In these galaxies, when we compare the weights of the two cosmic-ray electron (CRE) populations in model C2, the steep spectrum component is only of the order of 1$\sigma$–1.5$\sigma$ of the flatter spectrum component; thus, the extremely steep spectral index can be easily attributed to uncertainties in the data or flux measurements. Moreover, given that high-frequency observations (<7–8 GHz) can often be underestimated due to their lower S/N leading to a steep spectrum, the most likely cause of extremely steep spectra is due to the abovementioned effects.

3. For the majority of galaxies in our sample, the computed TF at 1.4 GHz (TF$_{1.4\,GHz}$) is <0.12 (Table 5) and is comparable to that of star-forming galaxies (Marvil et al. 2015).

4. CIGALE SED modeling resulted in the following range of astrophysical properties: dust luminosity, $\log_{10}(L_{dust})$ = 10.19–12.62, stellar mass, $\log_{10}(M_\star)$ = 9.71–11.78, IR SFR $\log_{10}$SFR$_{IR}$ = 0.29–2.95, optical AGN fraction = 1.3–99.0, radio-loudness parameter, $(R_{AGN})$ = 0.16–799.56, AGN PL slope, $\alpha_{AGN}$ = 0.61–1.03, star formation PL slope, $\alpha_{SF}$ = 0.64–0.90, $q_{IR}$ = 2.31–2.58, $\tau_V$ = 0.38–2.42 for our sample (Table 7). For all galaxies except IRAS 08572+3915, the $L_{dust}$, $M_\star$, SFR$_{IR}$, and AGN fractions are consistent with those typical for ULIRG-type galaxies (Małek et al. 2018).

5. We found that adding radio data to the SED modeling using CIGALE significantly increases the accuracy of the model fits. The median uncertainties on the derived $L_{dust}$ and SFR$_{IR}$ values are 0.02 and 0.03 dex, respectively. These are estimated to be than 1 order of magnitude better than previously reported (0.25 and 0.39 dex,

Table 7
CIGALE SED Fitting Results

| Name<br>(1) | $\log_{10}(L_{dust})$<br>$(L_\odot)$<br>(2) | $\log_{10}(M_\star)$<br>$(M_\odot)$<br>(3) | $\log_{10}(SFR_{IR})$<br>$(M_\odot$ yr$^{-1})$<br>(4) | AGN Fraction<br>(%)<br>(5) | Rad. Loud.<br>($R_{AGN}$)<br>(6) | $\alpha_{AGN}$<br>(7) | $\alpha_{SF}$<br>(8) | $q_{IR}$<br>(9) | $\tau_V$<br>(10) | $\chi^2$<br>(11) |
|---|---|---|---|---|---|---|---|---|---|---|
| IRAS F00183-7111 | 12.52 ± 0.02 | 10.90 ± 0.17 | 2.95 ± 0.02 | 49.98 ± 0.45 | 30.00 ± 0.01 | −0.83 ± 0.12 | −0.81 ± 0.17 | 2.37 ± 0.16 | 2.41 ± 0.01 | 2.4 |
| IRAS F03538−6432 | 12.48 ± 0.06 | 11.09 ± 0.12 | 2.90 ± 0.10 | 30.00 ± 0.21 | 9.93 ± 0.61 | −0.82 ± 0.07 | −0.81 ± 0.08 | 2.51 ± 0.09 | 1.61 ± 0.02 | 4.0 |
| IRAS 08572+3915 | 10.19 ± 0.05 | 10.32 ± 2.51 | 0.29 ± 0.06 | 99.00 ± 0.004 | 0.16 ± 0.01 | −0.61 ± 0.03 | −0.64 ± 0.04 | 2.53 ± 0.04 | 0.38 ± 0.12 | 3.5 |
| UGC 5101 | 11.88 ± 0.02 | 10.74 ± 0.06 | 2.00 ± 0.02 | 19.99 ± 0.001 | 5.00 ± 0.06 | −0.92 ± 0.12 | −0.75 ± 0.05 | 2.31 ± 0.05 | 2.30 ± 0.01 | 6.3 |
| IRAS 10565+2448 | 11.73 ± 0.02 | 10.56 ± 0.04 | 1.86 ± 0.03 | 3.00 ± 0.04 | 1.76 ± 0.75 | −0.75 ± 0.05 | −0.76 ± 0.05 | 2.39 ± 0.02 | 2.28 ± 0.001 | 4.8 |
| IRAS 12112+0305 | 11.68 ± 0.02 | 10.24 ± 0.06 | 1.90 ± 0.02 | 8.02 ± 0.18 | 5.2 ± 3.00 | −0.79 ± 0.03 | −0.75 ± 0.05 | 2.43 ± 0.02 | 2.19 ± 0.01 | 7.2 |
| UGC 08058 | 12.15 ± 0.02 | 12.22 ± 0.19 | 2.41 ± 0.03 | 40.00 ± 0.09 | 1.07 ± 0.26 | −0.70 ± 0.03 | −0.72 ± 0.07 | 2.48 ± 0.07 | 1.58 ± 0.01 | 4.3 |
| IRAS 13305-1739 | 12.09 ± 0.02 | 10.84 ± 0.05 | 2.23 ± 0.05 | 9.90 ± 2.00 | 15.44 ± 3.73 | −0.83 ± 0.13 | −0.83 ± 0.12 | 2.42 ± 0.11 | 1.64 ± 0.01 | 2.7 |
| UGC 8696 | 11.97 ± 0.02 | 10.56 ± 0.03 | 2.09 ± 0.04 | 2.11 ± 0.56 | 2.54 ± 1.36 | −0.75 ± 0.05 | −0.75 ± 0.05 | 2.42 ± 0.05 | 2.34 ± 0.02 | 3.3 |
| IRAS 14348-1447 | 12.07 ± 0.02 | 9.71 ± 0.08 | 2.45 ± 0.05 | 1.30 ± 0.46 | 2.92 ± 1.15 | −0.80 ± 0.08 | −0.70 ± 0.08 | 2.37 ± 0.07 | 2.19 ± 0.03 | 9.5 |
| IRAS 14394+5332 | 12.05 ± 0.02 | 10.50 ± 0.06 | 2.26 ± 0.03 | 4.95 ± 0.66 | 12.50 ± 4.29 | −0.75 ± 0.05 | −0.75 ± 0.05 | 2.35 ± 0.11 | 2.08 ± 0.04 | 3.6 |
| IRAS 17179+5444 | 12.18 ± 0.02 | 10.85 ± 0.04 | 2.29 ± 0.05 | 5.31 ± 1.20 | 199.77 ± 4.72 | −0.76 ± 0.05 | −0.75 ± 0.05 | 2.43 ± 0.14 | 2.35 ± 0.01 | 7.4 |
| IRAS F23529-2119 | 12.62 ± 0.02 | 11.53 ± 0.08 | 2.91 ± 0.02 | 30.54 ± 3.2 | 3.02 ± 0.23 | −0.81 ± 0.1 | −0.70 ± 0.15 | 2.46 ± 0.14 | 1.77 ± 0.003 | 2.2 |
| IRAS 23389+0300 | 11.72 ± 0.02 | 10.21 ± 0.43 | 1.98 ± 0.03 | 8.52 ± 0.87 | 799.56 ± 26.12 | −1.03 ± 0.09 | −0.90 ± 0.16 | 2.58 ± 0.10 | 2.34 ± 0.01 | 5.2 |

**Note.** Columns: (1) galaxy name, (2) stellar mass, (3) dust mass, (4) dust temperature, (5) instantaneous SFR, (6) AGN fraction, (7) slope of PL synchrotron emission, $\alpha$, (8) dust luminosity, (9) $q_{IR}$, (10) V-band optical depth, (11) reduced $\chi^2$ value for the best-fit model.

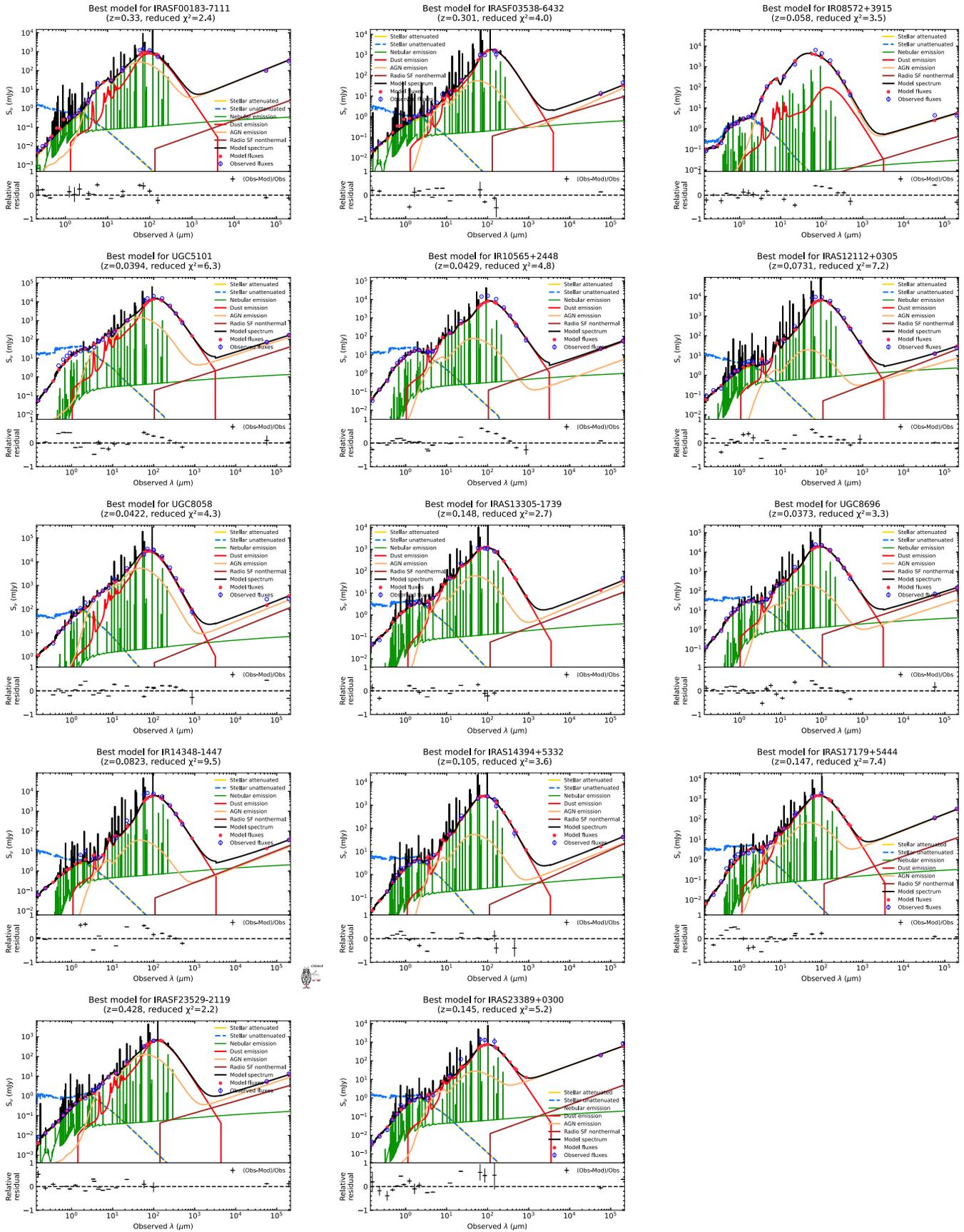

**Figure 2.** Best-fit UV-radio SEDs of our sample of 14 ULIRGs obtained using CIGALE.. The open and filled symbols represents the observed and modeled flux densities. The goodness of fit is estimated by the reduced $\chi^2$ shown at the top of each panel, along with the name and redshift of the galaxy. SEDs are well modeled in almost all cases, giving reasonable estimates of astrophysical properties.

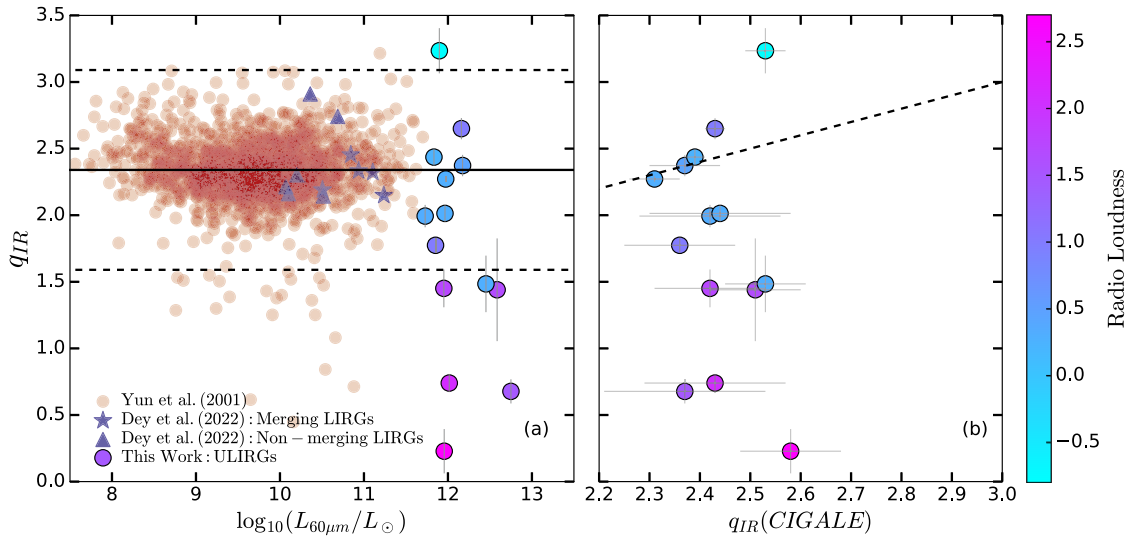

**Figure 3.** $q_{IR}$ estimated from Equation (1) vs. 60 μm dust luminosity. The solid line represents the mean, while the upper and lower dashed lines represent its 3σ bounds (panel (a)). Panel (b) shows the $q_{IR}$ estimated from Equation (1) against that from CIGALE. The color bar shows the decimal logarithmic radio-loudness. To help guide the eye, a dashed line showing the one-to-one line is plotted in panel (b), to visualize the correspondence between the two $q_{IR}$ estimation methods. All objects with radio-loudness ⩾10 are considered radio-loud.

respectively; Małek et al. 2018). A similar improvement in the measurement of $L_{dust}$ and $SFR_{IR}$ for star-forming galaxies was reported by Thorne et al. (2023) using the PROSPECT SED fitting tool, which included the radio module in the SED fitting code.

6. The most recent version of the CIGALE tool decomposes the total radio luminosity into AGN and star formation radio luminosities, allowing us to obtain $R_{AGN}$ and $\alpha_{AGN}$. Using the classical division between radio-loud and radio-quiet AGNs (Kellermann et al. 1989), six of our galaxies turn out to be radio-loud (IRAS F00183-7111, IRAS F03538-6442, IRAS 13305-1739, IRAS 14394 +5332, IRAS 17179+5444, and IRAS 23389+0300). Five out of six radio-loud ULIRGs have gigahertz peaked spectrum (GPS)-type AGN at their center (except for IRAS F03538-6442), which is the cause of radio-loudness. We also note that $\alpha_{AGN}$ is consistent with that of the canonical synchrotron emission spectral index.

## 6. Discussion

Below we discuss the global astrophysical properties for our entire sample of IR bright galaxies consisting of 11 LIRGs (studied in our previous paper; Dey et al. 2022) and 14 ULIRGs (this work). We performed a similar analysis on both samples of galaxies; hence, the results can be interpreted without any additional assumptions.

### 6.1. q$_{IR}$ Parameter

Within the framework of the *calorimetry* model of galaxies, the radio emission follows a tight empirical relation with the IR emission, denoted by $q_{IR}$ parameter (IR–radio correlation; Voelk 1989; Lacki et al. 2010). The origin of this relation lies within the physics of cosmic rays, magnetic fields, and interstellar dust, which are connected through star formation activity. Young OB stars heat their surrounding dust, which reemits the thermal continuum in the IR, and the supernova explosions of the same stars produce CREs that emit the radio continuum (Draine 2011; Klessen & Glover 2016). The galaxies are considered optically thick to radio and UV–IR emission in the sense that CREs lose all their energy via synchrotron radiation within the galaxy and that all UV photons emitted by young stars are absorbed by dust and then reemitted in the IR (Draine 2003). However, with the advent of the Spitzer and Hershel telescopes, it was noted that galaxies are optically thin to UV–IR emission instead, and a radio–FIR relation can be maintained with gas and magnetic field coupling in galaxies (Helou et al. 1988; Niklas et al. 1997; Lacki et al. 2010; Tabatabaei et al. 2017). Nevertheless, the radio–FIR correlation holds for four decades in IR luminosities, and for different galaxy classes (Yun et al. 2001; Magnelli et al. 2015), as well as within galaxies (Dumas et al. 2011; Tabatabaei et al. 2013).

The $q_{IR}$ parameter is given as (Yun et al. 2001)

$$q_{\mathrm{IR}} \equiv \log_{10}\left(\frac{\mathrm{FIR}}{3.75 \times 10^{12}\,\mathrm{W\,m^{-2}}}\right) - \log_{10}\left(\frac{S_{1.4\,\mathrm{GHz}}}{\mathrm{W\,m^{-2}\,Hz^{-1}}}\right) \quad (1)$$

where the FIR luminosity is computed using FIR ≡ $1.26 \times 10^{-14}(2.58\,S_{60\,\mu m} + S_{100\,\mu m})$ W m$^{-2}$, where $S_{60}$ and $S_{100\,\mu m}$ are the 60 and 100 μm band flux densities, in jansky, from IRAS and $S_{1.4\,GHz}$ is the 1.4 GHz radio luminosity.

Figure 3 shows the $q_{IR}$ estimated from Equation (1) and plotted against the monochromatic IR luminosity at 60 μm (panel (a)). The $L_{60\,\mu m}$ is derived as $4\pi\,d_L^2 \times S_{60\,\mu m}$ where $d_L$ is the luminosity distance. To obtain the luminosity distance, we use the cosmological calculator[7] (Wright 2006) with the Hubble constant $H_0 = 69.6$ km s$^{-1}$ Mpc$^{-1}$, $\Omega_M = 0.286$, and $\Omega_\lambda = 0.714$ (Bennett et al. 2014). Panel (b) shows the $q_{IR}$ estimated from CIGALE, $q_{IR}$(CIGALE), with the color bar showing the radio-loudness parameter. We note that $q_{IR}$(CIGALE) considers the dust continuum luminosity calculated based on the Bayesian analysis of the templates used in the rest-frame wavelength region of 8–1000 μm. In panel (a), the upper and lower dashed lines represent the 3σ bounds of the mean value with objects designated as *IR excess* and *radio excess* if they fall above and below the $3\sigma$ bounds, respectively. Most of our sample exhibits typical values of $q_{IR}$, indicative of radio emission from star formation processes alone, with six ULIRGs being radio excess objects, of which five show a kiloparsec-scale double-lobed radio galaxy at their center (IRAS F00183-7111, IRAS 13305-1739, IRAS 14394+5332, IRAS 17179+5444, IRAS 23389+0300), except for one galaxy, IRAS F03538-6432, where the cause of intense radio emission could be enhanced star formation, due to merging or a presence of a compact radio AGN. We also note that IRAS 08572+3915 is an IR excess object, due to the presence of heavily obscured AGN (Appendix C). Furthermore, the distribution of $q_{IR}$, with respect to $\log_{10} L_{60\mu m}/(L_\odot)$, gives comparable results to those obtained for other LIRGs by Yun et al. (2001) and Galvin et al. (2018). Here, LIRGs and ULIRGs (which are either nonmerging or merging/interacting) exhibit similar $q_{IR}$ values and are indistinguishable in this aspect. Furthermore, (Yun et al. 2001) obtained a mean value of $q_{IR}$ with $1\sigma$ scatter $= 2.34 \pm 0.26$ for a sample of 1809 galaxies that includes IR excess and radio excess objects in a small number (<2% of the total sample). Bell (2003) obtained a median value of $q_{IR}$ with $1\sigma$ scatter $= 2.64 \pm 0.26$ for a sample of 162 galaxies without AGN activity. For our sample of 25 galaxies using Equation 1, we get a mean with $1\sigma$ scatter $= 2.03 \pm 0.7$. Within the uncertainties, the $q_{IR}$ obtained for our sample is consistent with these studies.

---

[7] http://www.astro.ucla.edu/~wright/CosmoCalc.html

Panel (b) in Figure 3 shows the $q_{IR}$ estimated from Equation (1) plotted against the $q_{IR}$ obtained from UV-radio SED modeling using CIGALE ($q_{IR}$ (CIGALE)). The values of $q_{IR}$(CIGALE) for our sample of galaxies lie around a rather uniform value of $\sim 2.58$, in contrast to the values found via Equation (1), which show a range between $\sim 0.2$ and $\sim 3.2$. This is because CIGALE 2022.1 computes the $q_{IR}$ using the radio module where synchrotron spectral indices from star formation ($\alpha_{SF}$) and AGN ($\alpha_{AGN}$) and $q_{IR}$ are provided as an input. The radio module decomposes the total radio emission into that from star formation and AGN using $q_{IR}$ and $R_{AGN}$ as normalizing factors. Since we decompose the modeled star formation and AGN radio emission in our analysis, we obtain $q_{IR}$(CIGALE) only from the star-forming component, while the $q_{IR}$ estimated using Equation (1) considers the total radio luminosity irrespective of origin. That is why the $q_{IR}$(CIGALE) estimate is $\sim 2.5$, while the range is between $\sim 0.2$ and $\sim 3.2$ when $q_{IR}$ is estimated from Equation (1). We performed a Spearman's rank correlation test between $q_{IR}$ and $q_{IR}$(CIGALE) values, which measures the statistical dependence ($p$-value) and the strength of the correlation ($\rho$-value) between the two variables (Spearman 1904). A no-correlation null hypothesis is evaluated against a nonzero correlation hypothesis at a significance level of 0.01. The hypothesis of no correlation is rejected at a confidence level >99% for $p < 0.01$. The values of $q_{IR}$ and $q_{IR}$(CIGALE) of the ULIRGs sample are not correlated with $p = 0.55$ and $\rho = -0.17$. We show that the $q_{IR}$ values for these galaxies lie around 2.5 when the AGN contribution is taken into account.

### 6.2. Radio Spectral Indices and Spectral Curvature

The integrated radio spectra for our sample of ULIRGs show complex forms and are rarely described by single PLs; instead, they show multiple low-frequency bends and turnovers (Figure 1). Bends in the radio SED can occur not only due to FFA or SSA but also due to the Tsytovitch–Razin effect when the refractive index of the medium is less than unity (Israel & Mahoney 1990). The cutoff frequency for SSA to be effective is related to the radio surface brightness, which is given as peak flux/$\theta^2$, where the flux is in jansky, and $\theta$ is the size in arcsec (for unresolved sources, we take the synthesized beam size as the upper limit for the size). Typically, it is of the order of $10^{-2}$ for our sources. For the SSA mechanism to be important at 100 MHz for our sources, unreasonably high magnetic field strengths >1000 G are needed (Kellermann & Pauliny-Toth 1969). We further note that five ULIRGs in our sample are known to host GPS-type AGN and exhibit bent spectra (Appendix C). The estimated surface brightness for them is $\approx 10^{-3}$–500 when their very long baseline interferometry (VLBI) sizes and peak fluxes are considered. The corresponding cutoff frequencies range from 0.11–60 MHz. These estimated cutoff frequencies are smaller than the low-frequency turnovers in our cases; hence, SSA can be safely ruled out (Kellermann & Pauliny-Toth 1969).

The Tsytovitch–Razin turnover frequency is calculated as $20 \times N_e/B$(MHz), where $N_e$ is the electron density (per cubic centimeter) and $B$ is the magnetic field strength in microgauss (Ginzburg & Syrovatskii 1965). In ISM, for the typical CRE density $\sim 1$ (cm$^{-3}$) (Ferrière 2001) and the typical magnetic field strength of 50 $\mu$G (Crocker et al. 2010), the Tsytovitch–Razin turnover frequency turns out to be 0.4 MHz. This safely rules out the Tsytovitch–Razin effect as a cause of low-frequency turnover for our sources. The synchrotron spectral index for our sample of galaxies ranges between $\sim -0.6$ and $\sim -2.0$. The relatively flat spectra in galaxies, $\sim -0.6$, are comparable to the spectral index of synchrotron emission by electrons accelerated, due to diffusive shock acceleration with an energy index in the range of $-2$ to $-2.5$ in the shock fronts of supernova remnants (see Lacki et al. 2010, and the references therein). Steeper spectral slopes ranging from $-1.0$ to $-1.5$ can be explained by synchrotron/inverse-Compton cooling losses, as both are more efficient at higher frequencies (Hummel 1991; Marvil et al. 2015).

### 6.3. Thermal Emission

The thermal emission of ionized gas depends on the emission measure (EM), which is an integral of the electron density along the line of sight. EM is calculated by assuming that the emission originates from a volume of cylindrical geometry with constant temperature and electron density (Condon 1992). It is calculated from the optical depth, $\tau_\nu$, at the turnover frequency of

$$\tau_\nu = 3.28 \times 10^{-7} \left(\frac{T_e}{10^4 \text{ K}}\right) \left(\frac{\nu}{\text{GHz}}\right)^{-2.1} \left(\frac{\text{EM}}{\text{pc cm}^{-6}}\right) \quad (2)$$

where $T_e$ is the electron temperature of the H II emitting region and EM is the emission measure of depth $s$. $\tau_\nu$ is set to unity at the turnover frequency measured from our radio SED modeling. The EM is given by the integral of the electron density, $N_e$, along the line of sight of the H II region of depth $s$, assuming that $N_e^2$ does not change with $s$:

$$\frac{\text{EM}}{\text{pc cm}^{-6}} = \int_{\text{los}} \left(\frac{N_e}{\text{cm}^{-3}}\right)^2 d\left(\frac{s}{\text{pc}}\right). \quad (3)$$

**Table 8**
EM and Electron Density ($N_e$) Derived from the Value of the Free Parameters Obtained from the Best-fit Radio-only SED Model for Each Source

| Name | $EM_1$ ($10^6$ cm$^{-6}$ pc) | $N_{e,1}$ (cm$^{-3}$) | $EM_2$ ($10^6$ cm$^{-6}$ pc) | $N_{e,2}$ (cm$^{-3}$) |
|---|---|---|---|---|
| IRAS F00183-7111 | 0.57 | 2.98 | ⋯ | − |
| IRAS F03538-6432 | − | ⋯ | ⋯ | ⋯ |
| IRAS 08572+3915 | ⋯ | ⋯ | ⋯ | ⋯ |
| UGC 5101 | ⋯ | ⋯ | 2.56 | 153 |
| IRAS 10565+2448 | 0.30 | >37 | ⋯ | ⋯ |
| IRAS 12112+0305 | 0.18 | 26 | ⋯ | ⋯ |
| UGC 08058 | ⋯ | ⋯ | 28.11 | 693 |
| IRAS 13305+1739 | ⋯ | ⋯ | ⋯ | ⋯ |
| UGC 8696 | 0.06 | 16 | 210.36 | 939 |
| IRAS 14348-1447 | ⋯ | ⋯ | ⋯ | ⋯ |
| IRAS 14394+5332 | 0.02 | 6 | 145.46 | 537 |
| IRAS 17179+5444 | 1.15 | 83 | ⋯ | ⋯ |
| IRAS F23529-2119 | ⋯ | ⋯ | ⋯ | ⋯ |
| IRAS 23389+0300 | 0.28 | 21 | ⋯ | ⋯ |

**Note.** Sources without constrained EM and $N_e$ are marked by "⋯."

We compute the EMs for our sources using the turnover frequencies obtained from our modeling (Table 4) and assuming a typical electron temperature of $10^4$ K. Table 8 provides the EMs for our sample. Our EM values for our LIRG sample (Dey et al. 2022) and ULIRGs are consistent with those given for galaxies of these classes (Clemens et al. 2010; Galvin et al. 2018). We obtain mean ionized gas densities using EMs computed from our modeling (EM = $l$, where $l$ is the physical size of the source). For the majority of our sources, the physical sizes are calculated using angular sizes (=deconvolved major axis at full width at half-maximum) at 8.4 GHz from Condon et al. (1991) and distances computed following the methodology in Section 6.1. IRAS 10565+2448 was unresolved at 8.4 GHz (Condon et al. 1991), so we obtain a lower limit on the electron density because we use the resolution (= 0″.25) as a measure of angular size. For IRAS 14394+5332 and IRAS 23389+0300, we use the synthesized beam size from our radio imaging (Appendix A). For IRAS 17179+5444, we take a fitted major axis length as a measure of angular size (Myers et al. 2003). Since the observations rarely resolve the sources, we do not have a direct measure of the source sizes at turnover frequencies. These could, in principle, be larger than those at 8.4 GHz, due to cosmic-ray diffusion, but the synchrotron lifetimes are small at these frequencies, so the effect on the source size should be negligible; therefore, it is reasonable to use 8.4 GHz source sizes. For the Galvin et al. (2018) sources, we use the beam sizes at 9.0 GHz to measure the electron densities. The estimated electron densities are listed in Table 8 and are in accord with those found in the literature (Clemens et al. 2010).

### 6.4. TF

The TF measures the contribution of thermal FFE to the observed total radio continuum. It is an excellent indicator of current star formation in galaxies as it probes the H II region. The TF at 1.4 GHz ($TF_{1.4}$) was computed from the decomposed thermal and nonthermal flux estimated using the radio SED modeling. For our sample of ULIRGs, the TF is less than 0.11, except for the galaxy IRAS 08527+3915, whose value is ∼0.86 (Table 5). In Figure 4, we plot the $TF_{1.4}$ versus the IR luminosity from star formation processes alone (note that CIGALE gives dust luminosities from the AGN and star formation), with a color bar marking the bolometric AGN fraction. The TF values are consistent with those from previous studies (Marvil et al. 2015). We find that both samples of LIRGs and ULIRGs exhibit a similar range of TFs despite the expectation that ULIRGs should exhibit high TFs because they have high SFRs compared to LIRGs. However, a similar range could be observed, due to the absorption of ionizing photons by dust, therefore reducing the amount of gas-ionizing photons in LIRGs and ULIRGs (Valdés et al. 2005; Díaz-Santos et al. 2017). The ULIRG IRAS 08527+3915 exhibits about 2 orders of magnitude lower dust luminosity as compared to other ULIRGs (Table 7); therefore, the high TF shown by this source (=0.86) must be due to low absorption of ionizing photons by dust. However, the four ULIRGs showing low TF (<0.7%; Table 5) have dust luminosities comparable to other ULIRGs (∼$10^{12} L_\odot$; Table 7), which could be due to the large optical thickness of thermal ionized gas.

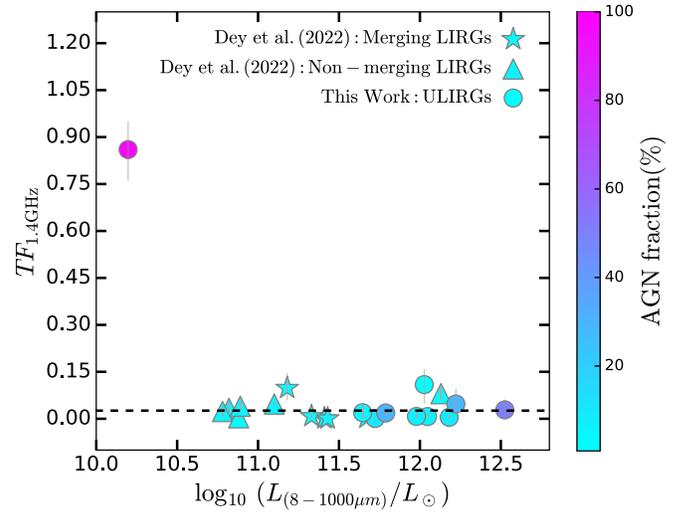

**Figure 4.** TF at 1.4 GHz plotted against the dust luminosity obtained from our CIGALE modeling. The dashed line indicates the median TF = 0.02 for our sample. The color bar represents the bolometric AGN fraction.

### 6.5. $TF_{1.4}$ versus $q_{IR}$

The $TF_{1.4}$ and $q_{IR}$ are good indicators of starburst timescale in star-forming galaxies. The TF is high in regions of recent star formation, due to increased thermal emission from accelerated free–free electrons in H II regions. The $q_{IR}$ also shows a similar nature, due to the increase in the IR emission and the deficit of nonthermal emission to the total radio continuum. Figure 5 compares TF with $q_{IR}$ with a color bar marking the timescale of star formation activity. In addition, we overplot the LIRGs from Dey et al. (2022) and the ULIRGs from Galvin et al. (2016) for comparison. As the starburst ages, we expect the TF and $q_{IR}$ to decrease, and the time-delayed synchrotron emission to set in. The LIRGs and ULIRGs do not follow the theoretical line by Marvil et al. (2015) well because many galaxies studied here are interacting, and therefore, undergo multiple bursts of star formation activity. No clear distinction between the young (cyan) and old (magenta) stellar populations is seen. The expectation of high TF in the case of ULIRGs undergoing young star formation (magenta) is not observed, possibly due to the absorption of ionizing photons by dust (Section 6.4).

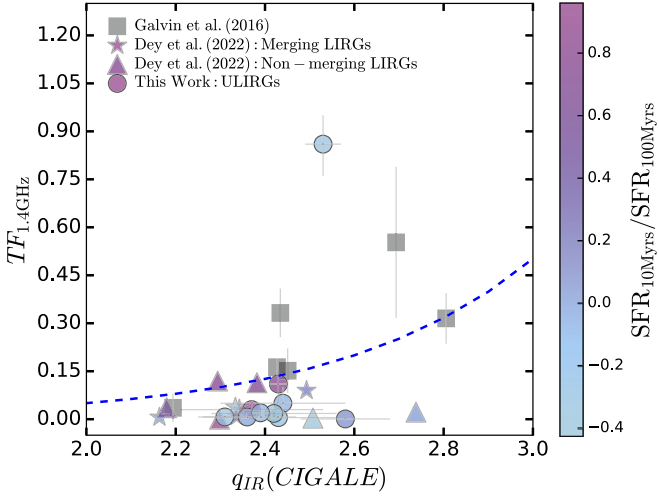

**Figure 5.** Distribution of TF with $q_{IR}$ (Equation (1)) for our sample of ULIRGs. The color bar indicates the ratio of $SFR_{IR}$ obtained for a time interval averaged over the last 10 and 100 Myr ago from the CIGALE analysis. The solid line marks the relation between TF and $q_{IR}$ given by TF = $1.7 \times 10^{q_{IR}-3.53}$ (Marvil et al. 2015).

### 6.6. SFR Calibration

One of the most important goals of our study is to calibrate radio continuum emission in order to derive the radio SFR, which is an excellent extinction-free SFR diagnostic. CIGALE modeling allows us to obtain $SFR_{IR}$ averaged over different time intervals, i.e., 10 and 100 Myr, enabling us to check the validity of the SFR calibration as a function of star formation history. Since our detailed radio SED modeling provides the decomposed nonthermal and thermal spectral luminosities, we obtain the total, nonthermal, and thermal SFRs following the SFR calibration relations given by Murphy et al. (2011). Thermal SFR, $SFR_\nu^{th}$ is derived using the Kroupa IMF (Kroupa 2001) assuming solar metalicity and continuous star formation:

$$\left(\frac{SFR_\nu^{th}}{M_\odot \, yr^{-1}}\right) = 4.6 \times 10^{-28} \left(\frac{T_e}{10^4 \, K}\right)^{-0.45} \left(\frac{\nu}{GHz}\right)^{0.1} \left(\frac{L_\nu^{th}}{erg \, s^{-1} \, Hz^{-1}}\right) \quad (4)$$

where $L_\nu^{th}$ is the thermal spectral luminosity and $T_e = 10^4$ K.

The nonthermal SFR, $SFR_\nu^{nth}$, is derived using calibration between the supernova rate and the SFR using the output of the Starburst99 model (Leitherer et al. 1999), an empirical relation between the supernova rate and the nonthermal spectral luminosity, $L_\nu^{nth}$, of the Milky Way (Tammann 1982; Condon & Yin 1990):

$$\left(\frac{SFR_\nu^{nth}}{M_\odot \, yr^{-1}}\right) = 6.64 \times 10^{-29} \left(\frac{\nu}{GHz}\right)^{\alpha^{nth}} \left(\frac{L_\nu^{nth}}{erg \, s^{-1} \, Hz^{-1}}\right) \quad (5)$$

where $\alpha^{nth} = \alpha = \alpha_{synch}$ is the synchrotron spectral index. For our calculations, we use the values obtained from the radio SED modeling (Table 4).

The total SFR, $SFR_\nu^{total}$, is obtained by combining Equations (4) and (5),

$$\left(\frac{SFR_\nu^{total}}{M_\odot \, yr^{-1}}\right) = 10^{-27} \left[2.18 \left(\frac{T_e}{10^4 \, K}\right)^{0.45} \left(\frac{\nu}{GHz}\right)^{-0.1} + 15.1 \left(\frac{\nu}{GHz}\right)^{-\alpha^{nth}}\right]^{-1} \left(\frac{L_\nu^{total}}{erg \, s^{-1} \, Hz^{-1}}\right). \quad (6)$$

This equation weights the total radio luminosity with the expected TF at a given frequency.

Figure 6 shows the $SFR_{IR}$ averaged over 10 and 100 Myr ago and total, nonthermal, and thermal radio SFRs obtained at 1.4 GHz (left panel) and 4.8 GHz (right panel). We obtain these radio SFRs from Equations (4)–(6) using thermal, nonthermal, and total radio fluxes from our radio SED modeling (see Section 4.1). We note that our CIGALE analysis used the Salpeter IMF for estimating the $SFR_{IR}$, while the radio SFRs are obtained using Kroupa's IMFs. This will lead to an offset between the IR and radio SFR estimates. To correct for this offset, we divide the radio SFRs by a constant factor of 0.67 (Madau & Dickinson 2014). The dashed line shows a one-to-one line to help guide the eye. Most of our sources fall on the line, with only a few exceptions. These sources have large radio luminosities, due to the presence of radio AGN (shown by large radio-loudness parameter), and hence, they do not follow a correspondence between IR and radio nonthermal and total luminosities.

To check for statistical dependence between the two independently estimated SFRs, we performed a Spearman's rank correlation test. We obtain a statistically significant correlation with $p \leqslant 0.01$ for the $SFR_{IR}$s averaged over 10 and 100 Myr, respectively, with total and nonthermal SFR at 1.4 GHz, respectively (Figures 6(a), (b), (d), (e); left panel), clearly rejecting the null hypothesis of no correlation with a confidence $\geqslant 99\%$. We obtained positive correlations between the two variables indicated by the positive $\rho$-values. Even though some galaxies deviate from the one-to-one line, due to high radio-loudness (shown by the color bar), the two independent SFR diagnostics are strongly correlated. This is quite puzzling as the typical lifetime of the CRE synchrotron at 1.4 GHz frequency is $\approx 3.4 \times 10^6$ yr (using Equation (18) of Lacki et al. 2010) for the assumed magnetic field strength of about 50 $\mu$G (Crocker et al. 2010). Therefore, a positive correlation between radio SFR and $SFR_{IR}$ for 10 Myr might be expected but not with $SFR_{IR}$ averaged over 100 Myr ago. However, we note that ULIRGs usually undergo multiple episodes of starburst activity (Farrah et al. 2003) during their lifetimes; therefore, an absence of a clear division between these two timescales with radio SFR is expected. Next, even though the thermal radio luminosity is due to star formation processes alone, we still do not recover a good correlation between the $SFR_{1.4GHz}^{th}$ and $SFR_{IR}$ ($p = 0.81$ and 0.91 for 10 and 100 Myr, respectively (Figures 6(c), (f), left panel). This could be due to the significant statistical uncertainty associated with measuring thermal luminosity at 1.4 GHz, where it is dominated by nonthermal processes (Condon 1992) and strengthens the view that thermal radio luminosity at 1.4 GHz is not a good tracer of recent star formation.

Similarly, we obtain a statistically significant correlation $p \leqslant 0.002$ for the $SFR_{IR}$ averaged over 10 Myr with total, nonthermal, and thermal radio SFR at 4.8 GHz (Figures 6(a)–(c), right panel), rejecting the null hypothesis with a confidence

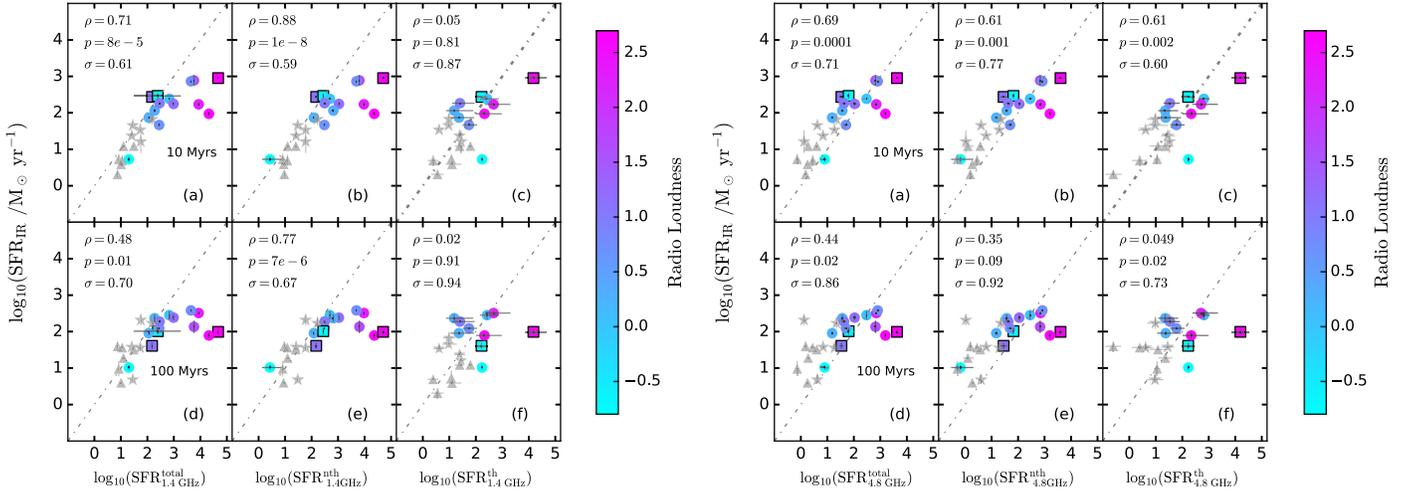

**Figure 6.** Comparison of the SFR$_{IR}$ obtained for two time intervals of 10 and 100 Myr ago from the CIGALE SED fitting with the radio total, nonthermal, and thermal SFRs, obtained at 1.4 GHz (left panel) and 4.8 GHz (right panel), respectively. The color bar shows the decimal logarithmic radio-loudness parameter. The circle and square show the burst age greater and smaller than 15 Myr, respectively, derived from the CIGALE modeling. To help guide the eye, a dashed line showing a one-to-one line in each panel is plotted.

>99.8%. We do not recover a statistically significant correlation for the SFR$_{IR}$ averaged over 100 Myr with total, nonthermal, and thermal radio SFRs at 4.8 GHz ($p \geqslant 0.02$; Figures 6(d)–(f), right panel). As the synchrotron lifetimes are about a factor of 2 smaller for electrons emitting at 4.8 GHz than at 1.4 GHz, similar results are expected when comparing SFR$_{IR}$ with radio SFRs obtained at these frequencies. As expected, the thermal radio SFR at 4.8 GHz shows a good correlation with SFR$_{IR}$ averaged over 10 Myr ago (Figure 6(c), right panel) compared to the thermal radio SFR at 1.4 GHz (Figure 6(c), left panel). If the optically thin condition holds, the thermal emission differs only slightly at 1.4 and 4.8 GHz (thermal spectral index is $-0.1$); therefore, our new result of a statistically significant correlation with thermal radio SFR at 4.8 GHz, with SFR$_{IR}$ averaged over 10 Myr ago compared to thermal radio SFR at 1.4 GHz, with SFR$_{IR}$ averaged over 10 Myr ago points to the fact that FFE is optically thick at 1.4 GHz, while it is optically thin at 4.8 GHz (also indicated by a low TF at 1.4 GHz; Section 6.4). Furthermore, we note that the SFR$_{IR}$ averaged over 100 Myr ago shows an indication of flattening with total and nonthermal radio SFRs compared to that averaged over 10 Myr, due to the low burst age obtained from our CIGALE analysis (shown with a circle and square with bust age greater and smaller than 15 Myr, respectively). This indicates that these galaxies have young star formation.

### 6.7. Comparison of Optical Morphological Differences and Their Relation to Attenuation, L$_{dust}$, SFRs, and Bolometric AGN Fractions

Our sample of LIRGs consisted of a total of 11 galaxies, with five merging galaxies and six isolated galaxies (Dey et al. 2022). Instead, all our ULIRGs are interacting, pre-merging, or in a post-merging stage (Appendix C). Figure 7(a) shows histograms of CIGALE estimated $V$-band optical depth, $\tau_V$ ($=0.921 \times A_V$, where $A_V$ is the attenuation at the $V$ band; Calzetti 2013). In our CIGALE modeling, we assume a single-component dust model by Calzetti et al. (2000). The two-sample Kolmogorov–Smirnov (K-S) test fails to reject the null hypothesis that the two data sets are of the same continuous distribution at a significance level of 0.01. Figure 7(b) shows a comparison between instantaneous SFR$_{IR}$ and $\tau_V$. Although they show about an order of magnitude higher SFRs than LIRGs, the ULIRGs exhibit the same range of $\tau_V$ as LIRGs. For our ULIRGs, this is a selection effect, as these are saturated in the $U$ band, creating a clustering at $\tau_V \approx 2$. On a separate note, we stress that the depth obtained from the mid-IR (MIR) spectra of the central kiloparsecs in some ULIRGs tends to be much larger than in LIRGs (see Stierwalt et al. 2013, 2014), and implies $\tau_V$ is much higher than what CIGALE can measure in the UV-optical wave bands because the nuclei are very thick. Figure 7(c) shows instantaneous SFR and $L_{dust}$ obtained from SED modeling using CIGALE. The dashed lines show the SFR obtained with the Kennicutt relation of SFR estimation from $L_{dust}$ following SFR$[M_\odot \text{ yr}^{-1}] = 4.5 \times 10^{-44} \times L_{dust}$ erg s$^{-1}$ (Kennicutt 1998). For both samples, the SFRs estimated from the CIGALE analysis show a good correspondence with those from the Kennicut's relation.

Regarding the bolometric AGN fraction, for LIRGs, it was limited to <10% (Dey et al. 2022), while for the majority of ULIRGs (11 out of 14), it is $\leqslant 30\%$ and reaches as high as 99% (Table 7). We performed a two-sample K-S test to check whether the bolometric AGN fractions depend on the optical morphological class. We divided our total sample into two subsamples: one consisting of isolated galaxies (six galaxies) and the other of merging or interacting galaxies (19 galaxies). The K-S test returned $p = 0.012$, which does not reject the null hypothesis that the two populations are drawn from the same continuous distribution at a 1% significance level. This supports previous conclusions that bolometric AGN fraction does not depend on the optical morphological class (e.g., Petric et al. 2011; Paspaliaris et al. 2021).

### 6.8. Comparison of Radio Spectral Indices with Redshift, SFR, Stellar Mass, Dust Mass, and sSFR

The radio continuum emission from star-forming galaxies is a combination of steep nonthermal synchrotron emission ($S_\nu \propto \nu^{-\alpha}$) and flat thermal FFE ($S_\nu \propto \nu^{-0.1}$), with the emission dominated by synchrotron emission at low frequencies ($\lesssim 10$ GHz; Condon 1992). For synchrotron emission from continuously injected CREs with a distribution of $N \propto E^{-\Gamma}$, the observed radio spectral index is $\alpha = (1-\Gamma)/2$, where $E$ and $\Gamma$ are the energy and particle energy index, respectively. When spectral aging is important, the synchrotron spectral index changes by $\Delta\alpha = 0.5$ at high frequencies as the particle distribution steepens to $N \propto E^{-\Gamma-1}$. Various absorption processes, escape of CREs, and adiabatic cooling of electrons further modify the shape of the radio continuum.

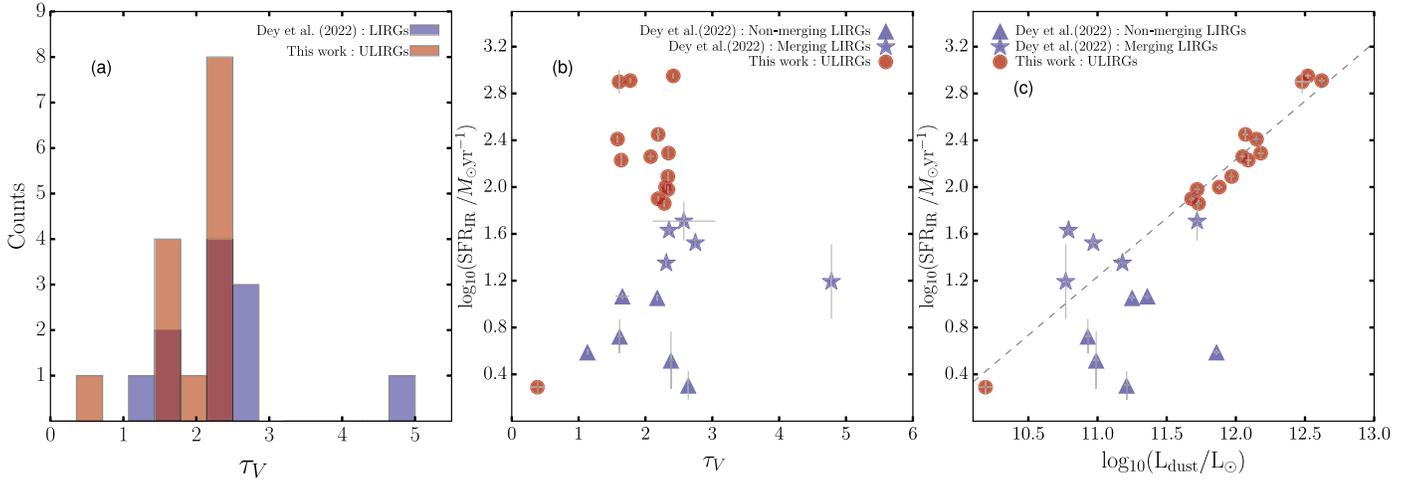

**Figure 7.** Histogram of $\tau_V$ (panel (a)), relation of $\tau_V$ and SFR$_{IR}$ (panel (b)), and relation of $L_{\rm dust}$ with SFR$_{IR}$ (panel (c)).

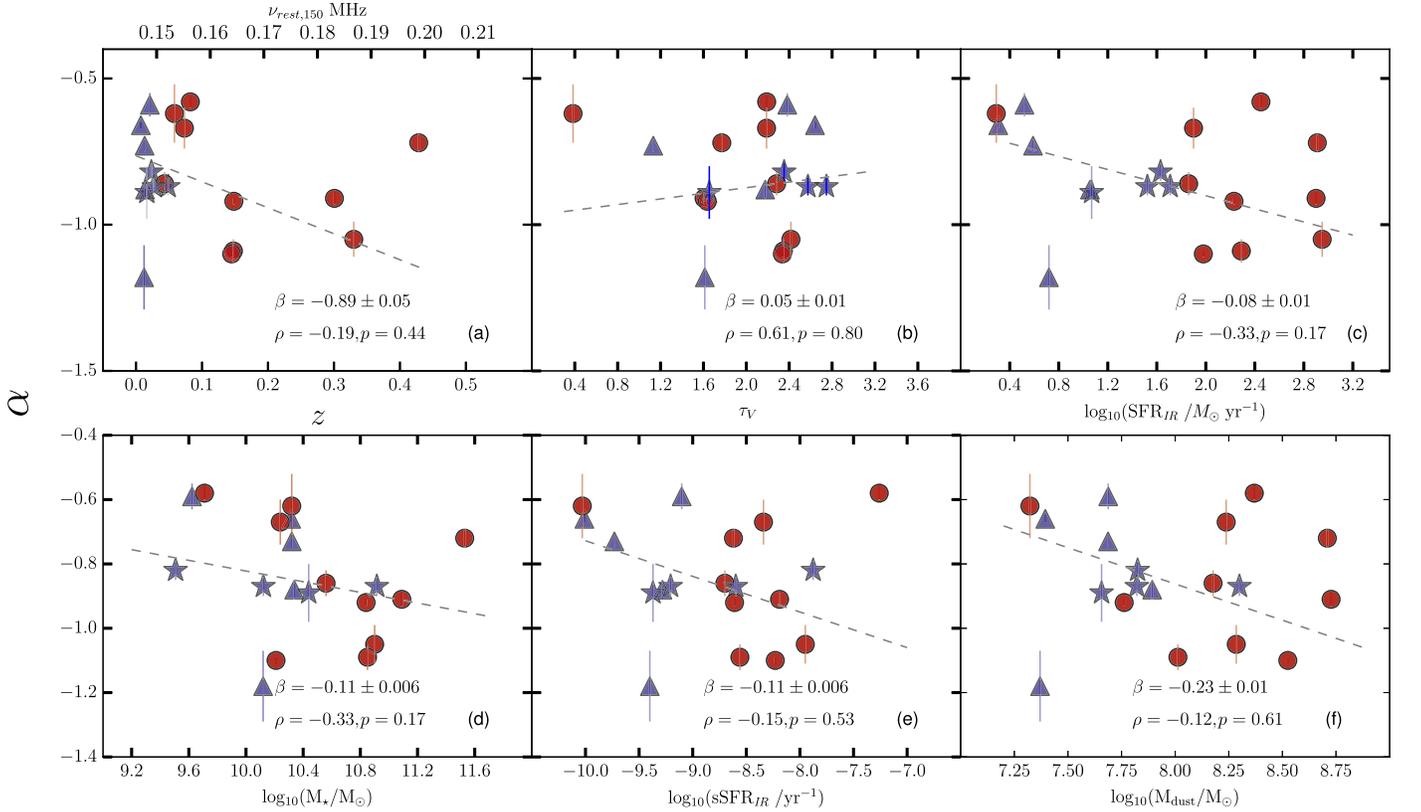

**Figure 8.** Variation of synchrotron spectral index, $\alpha$, obtained from radio SED modeling and redshift (panel (a)); SFR (panel (b)), stellar mass ($M_\star$; panel (c)), sSFR (panel (d)), and dust mass ($M_{\rm dust}$; panel (e)) for our truncated sample of 19 galaxies. The blue stars and triangles show the merging and nonmerging LIRGs, while the red circles represent the ULIRGs, all classified as merging or interacting galaxies. The slope of the weighted linear fit, along with 1$\sigma$ uncertainty and the $\rho$- and $p$-values from the application of Spearman's rank correlation test, are given inside each panel.

Figure 8 presents the scatter plots between the modeled synchrotron spectral index, $\alpha$ (Section 4.1) and redshift (panel (a)), SFR (panel (b)), stellar mass ($M_\star$; panel (c)), sSFR (panel (d)), and dust mass ($M_{\rm dust}$; panel (e)) obtained from the CIGALE modeling (Section 4.2) for our truncated sample of 19 galaxies (nine LIRGs and 10 ULIRGs). We chose these galaxies because, for them, the radio SEDs are best fitted by models where the nonthermal emission is due to a single population of accelerated particles. For this analysis, we exclude the six galaxies that best fit complex radio SEDs where the emission is due to two distinct populations of nonthermal particles. The weighted linear fit shown in Figure 8 is performed by minimizing $\chi^2$ and using the uncertainties of $\alpha$. The slopes of these linear fits, $\beta$, and 1$\sigma$ uncertainties are given in the right-bottom corner of each panel. Furthermore, we applied Spearman's rank correlation test. We do not find a statistically significant correlation between the analyzed quantities, as the null hypothesis of no correlation cannot be rejected.

Theoretically, from the single-zone model of CRE injection, cooling, and escape, one would expect a steepening of the radio spectra with redshift, due to increased inverse-Compton losses due to increased CMB density (Lacki et al. 2010; Klein et al. 2018). Our analysis does not reveal such evolution with redshift, possibly due to the low z-range ($z \leqslant 1$) and small sample size (panel (a); see also Magnelli et al. (2015) and An et al. (2021, 2023). We emphasize that we use the modeled spectral index in our analysis, where the flattening due to low frequencies (FFA) and at high frequencies (FFE) is taken care of. An et al. (2023) also noted the steepening of the observed two-point spectral index ($\alpha_{150\,\text{MHz}}^{610\,\text{MHz}}$) with $M_\star$ and instantaneous SFR$_{\text{IR}}$. Similarly, Heesen et al. (2022) also reported higher SFRs showing steeper spectra ($\alpha_{1.4\,\text{GHz}}^{150\,\text{MHz}}$) in their sample of 76 star-forming galaxies in the SFR range of $\sim 10^{-1}$–$10^{1} M_\odot$ yr$^{-1}$.

One expects galaxies with lower SFRs to show steeper spectra, due to the aging of CREs due to ceased acceleration. It is tempting to extend the Heesen et al. (2022) results to a higher SFR range as our sample consists of galaxies with an SFR range of $\sim 10^{0.4-3.0} M_\odot$ yr$^{-1}$. We find that $\alpha$ does not have a strong and statistically significant correlation with the SFR (large $p$-value and small $\rho$-value; panel (b)), though the normalization of linear fit seems to be a simple extrapolation from that obtained for lower SFR range galaxies (see Figure 5 (a) in Heesen et al. 2022). However, we cannot strictly compare our result with that of Heesen et al. (2022) because their spectral index is the observed one and not the clean synchrotron spectral index, used in our analysis. Moreover, their spectral index correlation with SFR cannot be due to the SFR, as the spectral index is flatter in star-forming regions in their galaxies (Appendix B of their manuscript). As shown globally, a flatter spectral index is expected with SFR surface density in galaxies (Tabatabaei et al. 2017).

Additionally, Heesen et al. (2022) show a stronger correlation of observed spectral index with stellar and dynamical mass (or rotation velocity) than with SFR, indicating that the spectral index–SFR correlation is secondary to galaxy size driving the rest of the correlations. The synchrotron spectral index–SFR correlation is not stronger than the spectral index $M_\star$ (Figure 8(c)) in our analysis, most likely due to the small sample size. Moreover, An et al. (2021) noted the steepening of the observed two-point spectral index ($\alpha_{1.3\,\text{GHz}}^{3.0\,\text{GHz}}$) with $M_\star$, while our analysis does not reveal such a trend (panel (c)). Galaxies with higher $M_\star$ are expected to be larger (Gürkan et al. 2018), causing CREs to take a longer time to escape from the galaxy; therefore, any steepening of $\alpha$ with $M_\star$ must be due to cooling losses.

An et al. (2021) noted the flattening of the observed two-point spectral index ($\alpha_{1.3\,\text{GHz}}^{3.0\,\text{GHz}}$) with instantaneous sSFR$_{\text{IR}}$ in the large SFR range $\gtrsim 10$. Similarly, Murphy et al. (2013) found an increase in the flattening of the observed radio spectral index ($\alpha_{1.4\,\text{GHz}}^{8.4\,\text{GHz}}$) with an increase in the sSFR. The sSFR is related to the compactness of the source (Elbaz et al. 2011). Murphy et al. (2013) interpreted it as due to compact starbursts hosting deeply embedded star formation that becomes more optically thick in the radio and IR. Our sample does not show such dependence because we use a modeled spectral index, where the flattening due to FFE is considered (panel (d)). We find no correlation between $\alpha$ and M$_{\text{dust}}$ for the sample (panel (e)), which is along the expectation that these galaxies retain the injection spectral index, as the CRE escape lifetimes are long because of their large sizes. An essential difference between our analysis and the studies mentioned above is that they rely on the observed spectral index obtained between two frequencies to determine the CRE dependence on other galaxy properties, while we have used a clean spectral index (from radio SED modeling), which takes into account the low-frequency flattening due to FFA and high-frequency flattening due to FFE. Moreover, a direct comparison between our analysis and the An et al. (2024) analysis is not possible as they do not assess their trends with any proper correlation or rank analysis as done here. Although our nonthermal spectral index is a true estimator of the underlying CRE population, we do not find any trends because of the large spread caused by the small sample size used in our analysis (19 galaxies).

## 7. Summary and Conclusion

In this study, we have simultaneously modeled densely sampled SEDs of 14 ULIRGs in the UV–IR–radio using the CIGALE SED modeling tool and radio-only wave bands ($\sim$54 MHz–30 GHz) using the UltraNest Bayesian inference tool. Including results from our previous study based on 11 LIRGs (Dey et al. 2022), we discuss the global insight into the evolution of astrophysical properties of IR bright galaxies. Our main conclusions are the following:

1. The sample consists primarily of Seyfert-type nuclei. Some galaxies show double or triple nuclei (e.g., IRAS 14394+5332), while none show large-scale radio jets. The radio emission is dominated by bright nuclei (e.g., Figure A1) and can be combined with IR to UV observations to model galaxy properties using CIGALE.

2. Our radio SED modeling using densely sampled SEDs of LIRGs and ULIRGs reveals that single PL forms rarely characterize their spectra. Instead, we observe many bends and turnovers. Some ULIRGs turn out to be radio-loud AGNs, due to the presence of GPS-type AGN in their centers; however, the bends in their radio SEDs are a result of FFA (also inferred in other studies, e.g., for IRAS F00183-7111 in Norris et al. 2012).

3. We note that 64% of ULIRGs and 63% of LIRGs show either one or two bends in their spectra. There is no indication that ULIRGs show more complicated radio spectra than LIRGs despite their optical morphologies displaying more signs of interaction and mergers. However, we note that our sample sizes are small.

4. The values of $q_{\text{IR}}$ and $q_{\text{IR}}$(CIGALE) for the ULIRGs sample are weakly correlated with $p = 0.55$. The large spread in $q_{\text{IR}}$ against $q_{\text{IR}}$(CIGALE) $\approx 2.5$ is due to the different definitions used for the calculation. Merging and nonmerging LIRGS have similar $q_{\text{IR}}$, but ULIRGs show larger spread and outliers. The $q_{\text{IR}}$ values are influenced by AGN, as indicated by their large radio-loudness.

5. ULIRGs exhibit a range of TFs comparable to LIRGs. This could be due to absorption of ionizing photons by dust, resulting in a decrease in the amount of ionizing gas present in ULIRGs. The EM for our LIRG and ULIRG samples are consistent with those found in the literature for galaxies of these classes.

6. Multiple studies have used CIGALE to model star-forming galaxies, LIRGs, and ULIRGs, but only a few included radio measurements in the SED fitting. Including radio measurements in the CIGALE modeling mainly better constrain the dust luminosity and SFR. Our results indicate that the uncertainties in the $L_{\mathrm{dust}}$ and SFR estimates are improved by more than an order of magnitude compared to those derived using UV–IR SED modeling alone.
7. We show that total and nonthermal radio luminosity at 1.4 GHz can be used as good estimators of recent SFRs for all LIRGs and ULIRGs that do not show large radio-loudness, indicative of radio-AN influence. A weaker but still significant correlation between radio SFR at 1.4 GHz and old (100 Myr) SFR based on SED modeling is observed, which indicates multiple episodes of starburst activity during their lifetimes.
8. We show that the total, nonthermal, and thermal radio luminosities at 4.8 GHz are also good tracers of recent SFRs for our sources. As expected, the thermal luminosity at 4.8 GHz turns out to be a better tracer of recent SFR than the thermal luminosity at 1.4 GHz. Our new result suggests that the FFE is optically thick at 1.4 GHz, while it becomes optically thin at 4.8 GHz.
9. Our results indicate that bolometric AGN fraction does not depend on the optical morphological class, consistent with those found in literature.
10. We compared the modeled radio nonthermal spectral indices obtained from radio SED modeling with redshift, stellar mass, dust mass, SFR, and sSFR. The nonthermal spectral indices are not statistically significantly correlated with the abovementioned properties.

## Acknowledgments

We thank the reviewer for the constructive comments on the manuscript, which improved the content and presentation. We thank Krzysztof Chyży for proofreading the manuscript and providing several valuable suggestions. We also thank Médéric Boquien for his suggestions. S.D. acknowledges the support of PROM PROGRAM (PPI/PRO/2019/1/00033/U/00001) "International scholarship exchange of PhD candidates and academic staff" (within the Operational Program Knowledge Education Development), financed by the Polish National Agency for Academic Exchange (NAWA) and cofinanced by the European Social Fund. K.M. has been supported by the National Science Center (UMO-2018/30/E/ST9/00082).

This research has made use of NED, which is operated by the Jet Propulsion Laboratory, California Institute of Technology, under contract with the National Aeronautics and Space Administration, the SIMBAD database, operated at CDS, Strasbourg, France 2000, "The SIMBAD astronomical database" (Wenger et al. 2000), the NASA/IPAC Infrared Science Archive, which is funded by the National Aeronautics and Space Administration and operated by the California Institute of Technology, and the VizieR catalog access tool, CDS, Strasbourg, France (Ochsenbein 1996).

*Facilities:* ASKAP, LOFAR, MWA, GMRT, VLA, Effelsberg, ATCA, GALEX, XMM, Skymapper, Swift, IRSA, Spitzer, WISE, Herschel, IRAS, AKARI.

*Software:* SKIRTOR (Stalevski et al. 2012, 2016), UltraNest (Buchner 2021), CIGALE (Noll et al. 2009; Boquien et al. 2019).

## Appendix A
## Summary of the Radio Observations Gathered for the Radio SED Modeling

Figure A1 provides an overlay of the 8.4 GHz radio contours of the PanStars I-band image for the galaxies IRAS 14394 +5332 and IRAS 23389+0300. Table A1 provides the integrated flux densities used in radio-only SED modeling. Figure A2 represents the corner plots for the check on the degeneracy of the model parameters.

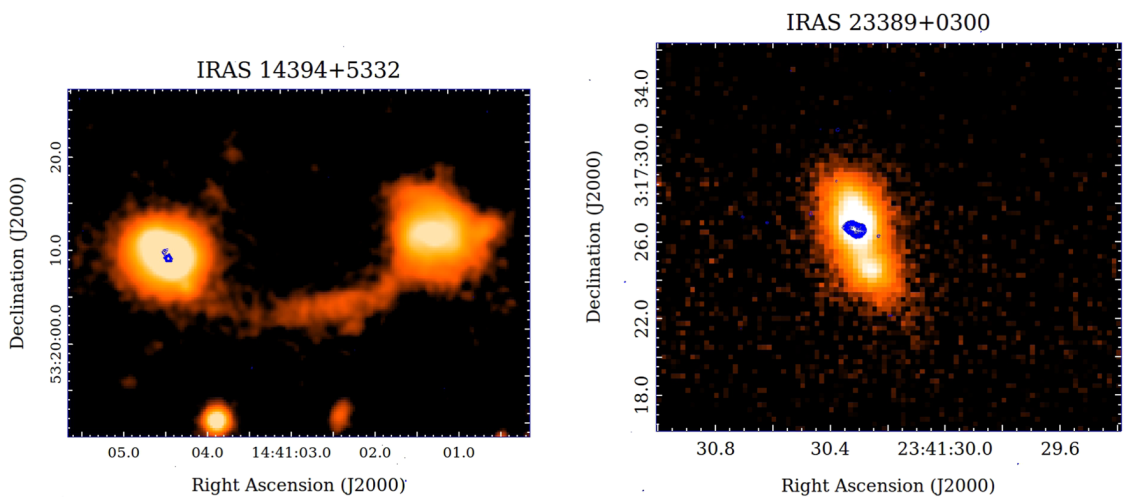

**Figure A1.** 8.4 GHz intensity contours (blue color) overlaid on a Pan-STARRS *I*-band image for the galaxies IRAS 14394+5332 (left panel) and IRAS 23389+0300 (right panel). Radio contours begin at $3\sigma$, where $\sigma$ is the background noise level and increase by factors of $(\sqrt{2})^n$ with $n = 0, 1, 2, 3....$ For the galaxies, IRAS 14394 +5332 and IRAS 23389+0300, the $\sigma$s are 65.96 and 66.33 $\mu$Jy beam$^{-1}$, respectively, and the synthesized beam sizes are $0\rlap{.}''26 \times 0\rlap{.}''24$ (position angle = 50°.80) and $0\rlap{.}''26 \times 0\rlap{.}''25$ (position angle = −62°.68), respectively.

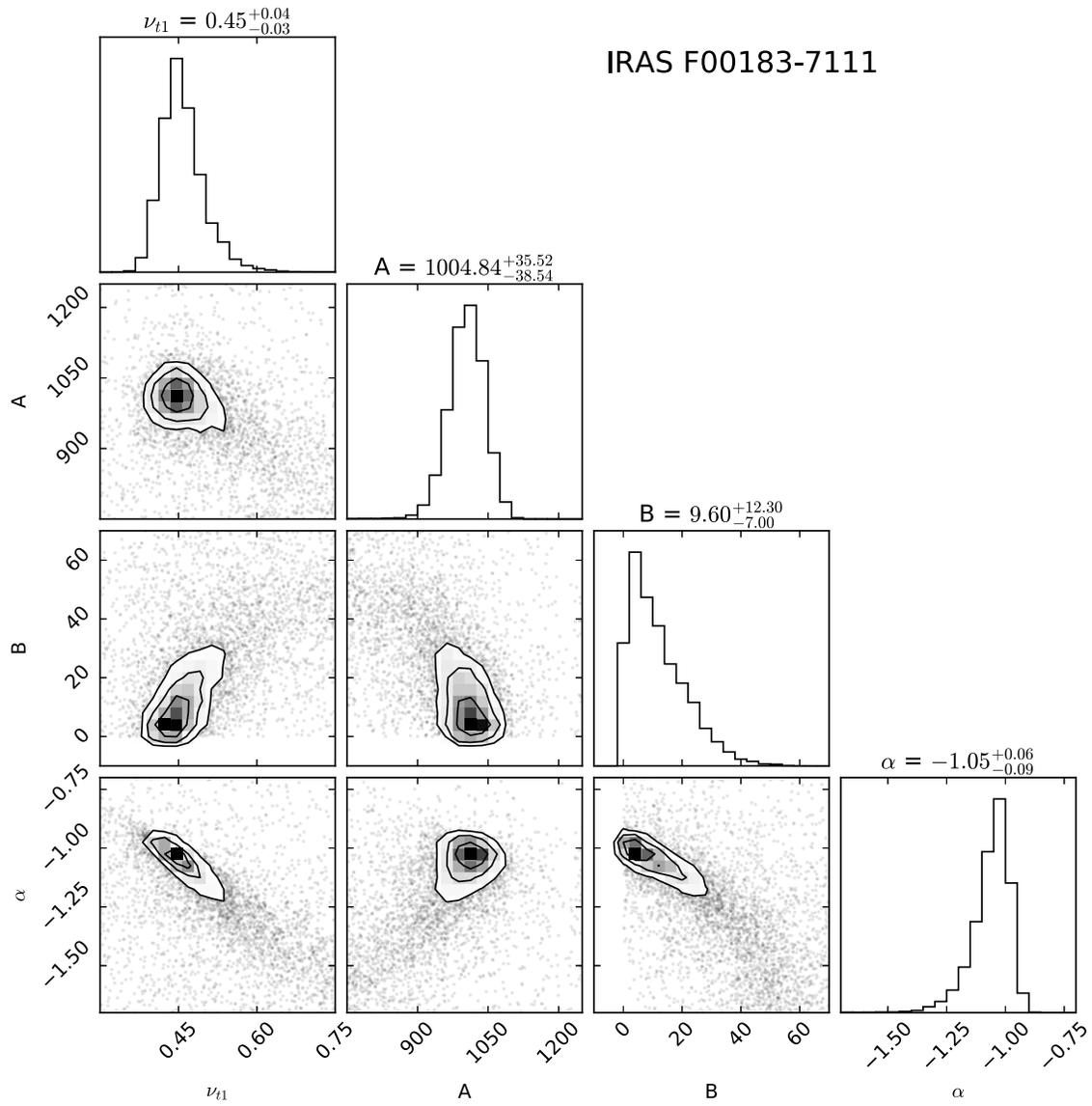

**Figure A2.** Corner plots showing the one and two-dimensional posterior probability distribution of the estimated parameters in the radio SED modeling for ULIRG IRAS F00183-7111. The complete figure set (14 images) is available in the online journal.

(The complete figure set (14 images) is available.)

**Table A1**
Integrated Radio Flux Densities Used for the SED Fitting

| Name | Obs. Freq. (GHz) | S (mJy) | Error (mJy) | References |
|---|---|---|---|---|
| (1) | (2) | (3) | (4) | (5) |
| IRAS F00183−7111 | 0.091 | 227.25 | 48.07 | (1) |
|  | 0.122 | 340.00 | 39.27 | (1) |
|  | 0.158 | 372.50 | 39.34 | (1) |
|  | 0.189 | 397.00 | 41.51 | (1) |
|  | 0.219 | 403.00 | 41.74 | (1) |
|  | 0.843 | 424.00 | 24.76 | (2) |
|  | 0.887 | 430.90 | 43.09 | (3) |
|  | 1.4[a] | 317.00[a] | 1.57[a] | (4) |
|  | 2.08 | 239.55 | 14.34 | (5) |
|  | 4.8[a] | 100.0[a] | 8.60[a] | (6) |
|  | 5.52 | 86.24 | 5.03 | (5) |
|  | 8.60 | 48.0 | 10.28 | (6) |
|  | 9.02 | 46.87 | 3.35 | (5) |
| IRAS F03538−6432 | 0.091 | 458.25 | 58.11 | (1) |
|  | 0.122 | 356.75 | 38.60 | (1) |
|  | 0.158 | 321.50 | 33.29 | (1) |
|  | 0.189 | 274.75 | 28.99 | (1) |
|  | 0.219 | 242.75 | 26.25 | (1) |
|  | 0.843 | 79.7 | 4.75 | (2) |
|  | 0.887 | 78.28 | 9.15 | (3) |
|  | 2.1 | 32.2 | 3.17 | (5) |
|  | 5.5[a] | 12.17[a] | 0.91[a] | (5) |
|  | 9.0 | 6.56 | 0.96 | (5) |
| IRAS 08572+3915 | 0.150 | 8.6 | 0.99 | (6) |
|  | 0.610 | 11.0 | 6.0 | (7) |
|  | 1.4[a] | 4.7[a] | 0.42[a] | (8) |
|  | 3.0 | 5.0 | 0.56 | (9) |
|  | 4.8[a] | 4.4[a] | 0.24[a] | (10) |
|  | 8.4 | 4.1 | 0.23 | (10) |
|  | 22.5 | 3.18 | 0.314 | (10) |
|  | 32.5 | 2.1 | 0.404 | (10) |
| UGC 5101 | 0.074 | 590.0 | 101.5 | (11) |
|  | 0.150 | 231.33 | 33.3 | (6) |
|  | 0.244 | 310.0 | 30.0 | (7) |
|  | 0.330 | 308.0 | 77.0 | (12) |
|  | 0.610 | 212.0 | 4.0 | (7) |
|  | 1.4[a] | 170.1[a] | 7.72[a] | (8) |
|  | 3.0 | 82.39 | 8.24 | (9) |
|  | 4.8[a] | 65.4[a] | 1.96[a] | (10) |
|  | 5.95 | 61.5 | 1.84 | (13) |
|  | 8.4 | 51.05 | 1.53 | (14) |
|  | 15.0 | 30.0 | 10.04 | (10) |
|  | 22.5 | 18.0 | 1.14 | (10) |
|  | 32.5 | 16.8 | 1.21 | (10) |
| IRAS 10565+2448 | 0.150 | 99.3 | 14.74 | (6) |
|  | 0.244 | 80.0 | 20 | (7) |
|  | 0.610 | 102.0 | 7.0 | (6) |
|  | 0.887 | 71.77 | 7.23 | (3) |
|  | 1.4[a] | 57.0[a] | 2.71[a] | (8) |
|  | 3.0 | 33.76 | 3.88 | (9) |
|  | 4.8[a] | 22.21[a] | 0.68[a] | (10) |
|  | 8.4 | 12.70 | 0.38 | (10) |
| IRAS 12112+0305 | 0.150 | 33.80 | 6.97 | (6) |
|  | 0.244 | 35.0 | 20.0 | (7) |
|  | 0.610 | 32.0 | 6.0 | (7) |
|  | 0.8875 | 33.306 | 4.409 | (3) |
|  | 1.4[a] | 24.0 | 1.39 | (8) |
|  | 3.0 | 17.0 | 1.73 | (9) |
|  | 4.8[a] | 12.44[a] | 0.7[a] | (15) |
|  | 8.4 | 8.6 | 0.26 | (16) |
|  | 14.9 | 5.75 | 0.32 | (15) |
|  | 22.5 | 5.7 | 0.33 | (10) |

Table A1
(Continued)

| Name | Obs. Freq. (GHz) | S (mJy) | Error (mJy) | References |
|---|---|---|---|---|
| (1) | (2) | (3) | (4) | (5) |
| UGC 8058 | 0.054 | 1336.78 | 11.08 | (16) |
| | 0.074 | 1000 | 123.69 | (11) |
| | 0.150 | 730.0 | 103.45 | (6) |
| | 0.244 | 590.0 | 20.0 | (7) |
| | 0.365 | 551.0 | 39.013 | (10) |
| | 0.610 | 403.0 | 20.0 | (7) |
| | 1.4[a] | 308.0 | 12.1 | (8) |
| | 3.0 | 293.66 | 29.37 | (9) |
| | 4.8[a] | 265.0[a] | 8.01[a] | (10) |
| | 5.95 | 313.0 | 9.39 | (13) |
| | 8.4 | 189.0 | 5.75 | (10) |
| | 10.63 | 190.0 | 11.16 | (17) |
| | 15.0 | 146.0 | 4.49 | (10) |
| | 22.5 | 137.0 | 4.57 | (10) |
| | 30.0 | 129.0 | 9.003 | (18) |
| | 32.5 | 88.0 | 4.38 | (13) |
| IRAS 13305-1739 | 0.091 | 377.0 | 51.28 | (1) |
| | 0.122 | 320.5 | 37.62 | (1) |
| | 0.158 | 258.00 | 27.68 | (1) |
| | 0.189 | 230.50 | 26.2 | (1) |
| | 0.219 | 209.75 | 22.43 | (1) |
| | 0.352 | 139.0 | 18.87 | (19) |
| | 0.887 | 68.07 | 8.61 | (3) |
| | 1.298 | 48.37 | 4.16 | (15) |
| | 1.4[a] | 48.0[a] | 3.6[a] | (8) |
| | 3.0 | 20.7 | 2.1 | (9) |
| | 8.44 | 6.3 | 0.34 | (15) |
| UGC 8696 | 0.054 | 488.55 | 19.64 | (16) |
| | 0.074 | 380.0 | 70.0 | (11) |
| | 0.150 | 454.4 | 64.55 | (6) |
| | 0.244 | 394.0 | 20.0 | (7) |
| | 0.365.0 | 335.0 | 37.0 | (10) |
| | 0.610.0 | 222.0 | 8.0 | (7) |
| | 1.4[a] | 144.7[a] | 6.69[a] | (8) |
| | 2.7 | 96.0 | 6.65 | (10) |
| | 3.0 | 88.4 | 8.86 | (9) |
| | 4.8[a] | 71.0[a] | 15.15[a] | (10) |
| | 8.4 | 43.75 | 2.17 | (10) |
| | 15.0 | 29.5 | 1.33 | (10) |
| | 22.5 | 14.3 | 0.52 | (10) |
| | 32.5 | 19.9 | 2.47 | (19) |
| IRAS 14348-1447 | 0.091 | 183.25 | 45.79 | (1) |
| | 0.122 | 105.15 | 21.91 | (1) |
| | 0.150 | 79.5 | 13.73 | (6) |
| | 0.158 | 90.2 | 15.12 | (1) |
| | 0.189 | 85.7 | 13.63 | (1) |
| | 0.219 | 92.0 | 12.22 | (1) |
| | 0.352 | 53.0 | 6.89 | (20) |
| | 0.610 | 30.0 | 6.0 | (6) |
| | 0.887 | 47.17 | 6.08 | (3) |
| | 1.4[a] | 35.9[a] | 1.61[a] | (8) |
| | 3.0 | 22.5 | 2.42 | (9) |
| | 8.4 | 10.8 | 1.05 | (10) |
| | 22.5 | 4.5 | 0.51 | (10) |
| IRAS 14394+5332 | 0.054 | 271.41 | 6.56 | (16) |
| | 0.150 | 230.0 | 23.0 | (15) |
| | 0.150 | 224.0 | 33.0 | (6) |
| | 0.327 | 142.0 | 35.11 | (15) |
| | 1.4[a] | 42.0[a] | 2.45[a] | (8) |
| | 3.0 | 23.72 | 2.38 | (9) |
| | 8.5 | 13.7 | 0.411 | (21) |
| | 15.0 | 5.6 | 0.034 | (15) |

Table A1
(Continued)

| Name | Obs. Freq. (GHz) | S (mJy) | Error (mJy) | References |
|---|---|---|---|---|
| (1) | (2) | (3) | (4) | (5) |
| IRAS 17179+5444 | 0.150 | 177.3 | 25.4 | (6) |
|  | 0.365 | 510.0 | 20.5 | (15) |
|  | 1.4[a] | 332.0[a] | 14.11[a] | (8) |
|  | 3.0 | 179.0 | 17.91 | (9) |
|  | 4.85[a] | 117.88[a] | 6.96[a] | (15) |
|  | 8.4 | 65.6 | 6.0 | (22) |
|  | 14.9 | 27.0 | 1.77 | (15) |
| IRAS F23529−2119 | 0.091 | 236.0 | 33.29 | (1) |
|  | 0.122 | 115.25 | 14.77 | (1) |
|  | 0.158 | 61.67 | 7.88 | (1) |
|  | 0.189 | 47.95 | 5.78 | (1) |
|  | 0.219 | 42.42 | 4.99 | (1) |
|  | 0.887 | 22.78 | 3.16 | (3) |
|  | 1.4[a] | 16.2[a] | 0.85[a] | (8) |
|  | 2.1 | 9.66 | 2.41 | (4) |
|  | 5.5[a] | 4.77[a] | 0.47[a] | (4) |
|  | 9.0 | 3.01 | 0.685 | (4) |
| IRAS 23389+0300 | 0.074 | 1751.0 | 222.17 | (11) |
|  | 0.150 | 2608.2 | 368.91 | (7) |
|  | 0.178 | 3400.0 | 892.46 | (15) |
|  | 0.365 | 2650.0 | 67.79 | (15) |
|  | 0.408 | 2225.48 | 356.3 | (15) |
|  | 0.653 | 1900.0 | 152.0 | (23) |
|  | 0.887 | 1330.24 | 162.68 | (3) |
|  | 1.298 | 845.0 | 72.68 | (15) |
|  | 1.4[a] | 863.0[a] | 36.69[a] | (8) |
|  | 3.0 | 390.0 | 39.05 | (9) |
|  | 4.85[a] | 199.0[a] | 16.15[a] | (24) |
|  | 6.0 | 194.0 | 1.94 | (21) |
|  | 8.4 | 95.0 | 3.5 | (21) |
|  | 15.0 | 41.0 | 4.28 | (15) |

**Notes.** Columns: (1) source name; (2) observing frequency; (3) integrated flux density of the source; (4) the error in the flux density including the absolute calibration uncertainty (Section 3.1); (5) References: (1) GLEAM (Wayth et al. 2015), (2) SUMSS (Bock et al. 1999), (3) ASKAP (Hale et al. 2021), (4) ATCA (Galvin et al. 2016), (5) McConnell et al. (2012), (6) TGSS ADR1 (Intema et al. 2017), (7) Clemens et al. (2010), (8) NVSS (Condon et al. 1998), (9) VLASS (Gordon et al. 2021), (10) Clemens et al. (2008), (11) VLSSr (Lane et al. 2014), (12) Rengelink et al. (1997), (13) Leroy et al. (2011), (14) Vardoulaki et al. (2015), (15) Nandi et al. (2021), (16) LoLLS (de Gasperin et al. 2023), (17) Bicay et al. (1995), (18) Lowe et al. (2007), (19) Barcos-Muñoz et al. (2017), (20) De Breuck et al. (2002), (21) reanalyzed VLA (https://data.nrao.edu/portal/#/), (22) Myers et al. (2003), (23) Wright & Otrupcek (1990), (24) Griffith et al. (1995).
[a] Used in CIGALE modeling.

# Appendix B
# Summary of the UV–IR Observations Gathered for SED Modeling via CIGALE

Table B1 lists the flux densities used in CIGALE modeling.

**Table B1**
Integrated UV–IR Flux Densities Used for the SED Fitting

| Name | Instrument | Passband | S (mJy) | Error (mJy) | Integration Area ($a \times b$) | References |
|---|---|---|---|---|---|---|
| (1) | (2) | (3) | (4) | (5) | (6) | (7) |
| IRAS F00183-7111 | GALEX | FUV | 0.012 | 0.004 | $7\rlap.{''}3 \times 7\rlap.{''}3$ | (1) |
|  |  | NUV | 0.012 | 0.001 | $3\rlap.{''}5 \times 3\rlap.{''}5$ | (1) |
|  | 2MASS | $J$ | 0.35 | 0.09 | $4'' \times 4''$ | (2) |
|  |  | $H$ | 0.41 | 0.13 | $4'' \times 4''$ | (2) |
|  |  | $K_s$ | 0.81 | 0.19 | $4'' \times 4''$ | (2) |
|  | Spitzer | IRAC1 | 2.9 | 0.3 | $12'' \times 12''$ | (3) |
|  |  | IRAC3 | 22.0 | 2.0 | $12'' \times 12''$ | (3) |
|  | WISE | W1 | 1.91 | 00.01 | $8\rlap.{''}25 \times 8\rlap.{''}25$ | (4) |
|  |  | W2 | 7.51 | 0.04 | $8\rlap.{''}25 \times 8\rlap.{''}25$ | (4) |
|  |  | W3 | 29.0 | 0.21 | $8\rlap.{''}25 \times 8\rlap.{''}25$ | (4) |
|  |  | W4 | 97.1 | 1.25 | $16\rlap.{''}5 \times 16\rlap.{''}5$ | (4) |
|  | IRAS | IRAS2 | 133.0 | 10.2 | $>0\rlap.{'}8 \times >0\rlap.{'}8$[a] | (6) |
|  |  | IRAS3 | 1200 | 83.7 | $>0\rlap.{'}8 \times >0\rlap.{'}8$[a] | (6) |
|  |  | IRAS4 | 1190 | 119.0 | $>0\rlap.{'}8 \times >0\rlap.{'}8$[a] | (6) |
|  | MIPS | MIPS2 | 1500 | 225.0 | $8'' \times 8''$ | (3) |
|  |  | MIPS3 | 540.0 | 80.0 | $8'' \times 8''$ | (3) |
| IRAS 03538-6432 | GALEX | FUV | 0.027 | 0.004 | $3\rlap.{''}6 \times 3\rlap.{''}6$ | (1) |
|  |  | NUV | 0.0052 | 0.004 | $3\rlap.{''}5 \times 3\rlap.{''}5$ | (1) |
|  | SMSS | $g$ | 0.19 | 0.02 | $2\rlap.{''}75 \times 2\rlap.{''}75$ | (7) |
|  |  | $r$ | 0.28 | 0.04 | $2\rlap.{''}75 \times 2\rlap.{''}75$ | (7) |
|  |  | $i$ | 0.20 | 0.04 | $2\rlap.{''}75 \times 2\rlap.{''}75$ | (7) |
|  |  | $z$ | 0.36 | 0.04 | $2\rlap.{''}75 \times 2\rlap.{''}75$ | (7) |
|  | 2MASS | $J$ | 0.55 | 0.05 | $4'' \times 4''$ | (2) |
|  |  | $H$ | 1.01 | 0.21 | $4'' \times 4''$ | (2) |
|  |  | $K_s$ | 1.27 | 0.05 | $4'' \times 4''$ | (2) |
|  | WISE | W3 | 15.3 | 0.127 | $8\rlap.{''}25 \times 8\rlap.{''}25$ | (4) |
|  |  | W4 | 57.6 | 0.903 | $16\rlap.{''}5 \times 16\rlap.{''}5$ | (4) |
|  | Spitzer | IRAC1 | 1.8 | 0.02 | $5\rlap.{''}8 \times 5\rlap.{''}8$ | (8) |
|  |  | IRAC2 | 2.72 | 0.02 | $5\rlap.{''}8 \times 5\rlap.{''}8$ | (8) |
|  |  | IRAC3 | 4.8 | 0.07 | $5\rlap.{''}8 \times 5\rlap.{''}8$ | (8) |
|  |  | IRAC4 | 12.0 | 0.1 | $5\rlap.{''}8 \times 5\rlap.{''}8$ | (8) |
|  | AKARI | N60 | 1020 | 350.0 | $40'' \times 50''$[b] | (9) |
|  |  | WIDE-S | 986.0 | 61.0 | $40'' \times 50''$[b] | (9) |
|  |  | WIDE-L | 1540 | 150.0 | $70'' \times 90''$[b] | (9) |
|  |  | N160 | 1050 | 420.0 | $70'' \times 90''$[b] | (9) |
| IR08572+3915 | GALEX | FUV | 0.085 | 0.01 | $3\rlap.{''}5 \times 3\rlap.{''}5$ | (1) |
|  |  | NUV | 0.128 | 0.008 | $3\rlap.{''}5 \times 3\rlap.{''}5$ | (1) |
|  | SDSS | $u'$ | 0.20 | 0.02 | $5\rlap.{''}87 \times 5\rlap.{''}87$ | (10) |
|  |  | $g'$ | 0.87 | 0.03 | $8.21'' \times 8\rlap.{''}21$ | (10) |
|  |  | $r'$ | 1.33 | 0.05 | $7\rlap.{''}71 \times 7\rlap.{''}71$ | (10) |
|  |  | $i'$ | 1.70 | 0.07 | $7\rlap.{''}11 \times 7\rlap.{''}11$ | (10) |
|  |  | $z'$ | 1.85 | 0.10 | $5\rlap.{''}46 \times 5\rlap.{''}46$ | (10) |
|  | 2MASS | $J$ | 3.10 | 0.47 | $14'' \times 14''$ | (2) |
|  |  | $H$ | 3.74 | 0.52 | $14'' \times 14''$ | (2) |
|  |  | $K_s$ | 4.42 | 0.53 | $14'' \times 14''$ | (2) |
|  | WISE | W1 | 23.9 | 2.21 | $8\rlap.{''}25 \times 8\rlap.{''}25$ | (4) |
|  |  | W2 | 111.0 | 10.0 | $8\rlap.{''}25 \times 8\rlap.{''}25$ | (4) |
|  |  | W3 | 325.0 | 29.9 | $8\rlap.{''}25 \times 8\rlap.{''}25$ | (4) |
|  |  | W4 | 1290.0 | 119.0 | $16\rlap.{''}5 \times 16\rlap.{''}5$ | (4) |
|  | Herschel-PACS | Blue | 6398.0 | 282.0 | $35'' \times 35''$ | (11) |
|  |  | Green | 4452.0 | 194.0 | $35'' \times 35''$ | (11) |
|  |  | Red | 1997.0 | 84.0 | $35'' \times 35''$ | (11) |
|  | Herschel-SPIRE | PSW | 521.0 | 31.0 | $45'' \times 45''$ | (11) |
|  |  | PMW | 188.0 | 14.0 | $45'' \times 45''$ | (11) |
|  |  | PLW | 42.0 | 7.0 | $45'' \times 45''$ | (11) |
| UGC 5101 | GALEX | FUV | 0.056 | 0.002 | $3\rlap.{''}5 \times 3\rlap.{''}5$ | (1) |
|  |  | NUV | 0.19 | 0.007 | $3\rlap.{''}5 \times 3\rlap.{''}5$ | (1) |
|  | SDSS | $u'$ | 0.087 | 0.01 | $13\rlap.{''}37 \times 13\rlap.{''}37$ | (10) |
|  |  | $g'$ | 4.02 | 0.01 | $13\rlap.{''}37 \times 13\rlap.{''}37$ | (10) |
|  |  | $r'$ | 8.40 | 0.002 | $13\rlap.{''}37 \times 13\rlap.{''}37$ | (10) |
|  |  | $i'$ | 13.20 | 0.026 | $13\rlap.{''}37 \times 13\rlap.{''}37$ | (10) |
|  |  | $z'$ | 19.41 | 0.066 | $13\rlap.{''}37 \times 13\rlap.{''}37$ | (10) |

Table B1
(Continued)

| Name | Instrument | Passband | S (mJy) | Error (mJy) | Integration Area ($a \times b$) | References |
|---|---|---|---|---|---|---|
| (1) | (2) | (3) | (4) | (5) | (6) | (7) |
| | 2MASS | $J$ | 15.7 | 0.31 | $23\rlap.{''}4 \times 14\rlap.{''}5$ | (2) |
| | | $H$ | 23.6 | 0.46 | $23\rlap.{''}4 \times 14\rlap.{''}5$ | (2) |
| | | $K_s$ | 31.6 | 0.59 | $23\rlap.{''}4 \times 14\rlap.{''}5$ | (2) |
| | WISE | W1 | 31.7 | 0.17 | $22'' \times 22''$ | (4) |
| | | W2 | 77.8 | 0.43 | $22'' \times 22''$ | (4) |
| | | W3 | 156.0 | 0.86 | $22'' \times 22''$ | (4) |
| | | W4 | 566.0 | 3.13 | $22'' \times 22''$ | (4) |
| | IRAS | IRAS2 | 1030 | 15.0 | $>0\rlap.{'}8 \times >0\rlap.{'}8^a$ | (6) |
| | | IRAS3 | 11,500 | 808.0 | $>0\rlap.{'}8 \times >0\rlap.{'}8^a$ | (6) |
| | | IRAS4 | 20,200 | 1010 | $>0\rlap.{'}8 \times >0\rlap.{'}8^a$ | (6) |
| | AKARI | N60 | 11,200 | 700.0 | $40'' \times 50''^b$ | (9) |
| | | WIDE-S | 14,600 | 1000 | $40'' \times 50''^b$ | (9) |
| | | WIDE-L | 15,600 | 1000 | $70'' \times 90''^b$ | (9) |
| | | N160 | 11,600 | 2700 | $70'' \times 90''^b$ | (9) |
| IR 10565+2448 | GALEX | FUV | 0.032 | 0.002 | $3\rlap.{''}5 \times 3\rlap.{''}5$ | (1) |
| | | NUV | 0.129 | 0.003 | $3\rlap.{''}5 \times 3\rlap.{''}5$ | (1) |
| | SDSS | $u'$ | 0.393 | 0.014 | $5\rlap.{''}87 \times 5\rlap.{''}87$ | (10) |
| | | $g'$ | 1.72 | 0.032 | $8\rlap.{''}21 \times 8\rlap.{''}21$ | (10) |
| | | $r'$ | 3.525 | 0.054 | $7\rlap.{''}71 \times 7\rlap.{''}71$ | (10) |
| | | $i'$ | 5.63 | 0.076 | $7\rlap.{''}11 \times 7\rlap.{''}11$ | (10) |
| | | $z'$ | 7.49 | 0.088 | $5\rlap.{''}46 \times 5\rlap.{''}46$ | (10) |
| | 2MASS | $J$ | 11.9 | 0.321 | $17\rlap.{''}8 \times 16.7''$ | (2) |
| | | $H$ | 16.7 | 0.53 | $17\rlap.{''}8 \times 16.7''$ | (2) |
| | | $K_s$ | 17.9 | 0.638 | $17\rlap.{''}8 \times 16.7''$ | (2) |
| | WISE | W1 | 14.6 | 0.081 | $22'' \times 22''$ | (3) |
| | | W2 | 15.6 | 0.10 | $22'' \times 22''$ | (3) |
| | | W3 | 155.0 | 0.854 | $22'' \times 22''$ | (3) |
| | | W4 | 655.0 | 3.62 | $22'' \times 22''$ | (3) |
| | IRAS | IRAS1 | 217.0 | 32.6 | $>0\rlap.{'}8 \times >0\rlap.{'}8^a$ | (6) |
| | | IRAS2 | 1140 | 13.0 | $>0\rlap.{'}8 \times >0\rlap.{'}8^a$ | (6) |
| | | IRAS3 | 12,100 | 606.0 | $>0\rlap.{'}8 \times >0\rlap.{'}8^a$ | (6) |
| | | IRAS4 | 15,100 | 757.0 | $>0\rlap.{'}8 \times >0\rlap.{'}8^a$ | (6) |
| | Herschel-PACS | Blue | 14,300 | 640.0 | $35'' \times 35''$ | (11) |
| | | Green | 15,800 | 710.0 | $35'' \times 35''$ | (11) |
| | | Red | 10,500 | 460.0 | $35'' \times 35''$ | (11) |
| | Herschel-SPIRE | PSW | 3640 | 215.0 | $45'' \times 45''$ | (11) |
| | | PMW | 1340 | 78.0 | $45'' \times 45''$ | (11) |
| | | PLW | 381.0 | 22.0 | $45'' \times 45''$ | (11) |
| IRAS 12112+0305 | GALEX | FUV | 0.09 | 0.002 | $3\rlap.{''}5 \times 3\rlap.{''}5$ | (1) |
| | | NUV | 0.17 | 0.002 | $3\rlap.{''}5 \times 3\rlap.{''}5$ | (1) |
| | SDSS | $u'$ | 0.21 | 0.02 | $8\rlap.{''}09 \times 8\rlap.{''}09$ | (10) |
| | | $g'$ | 0.62 | 0.02 | $8\rlap.{''}09 \times 8\rlap.{''}09$ | (10) |
| | | $r'$ | 1.03 | 0.02 | $8\rlap.{''}09 \times 8\rlap.{''}09$ | (10) |
| | | $i'$ | 1.64 | 0.03 | $8\rlap.{''}09 \times 8\rlap.{''}09$ | (10) |
| | | $z'$ | 1.77 | 0.08 | $8\rlap.{''}09 \times 8\rlap.{''}09$ | (10) |
| | 2MASS | $J$ | 3.94 | 0.36 | $18'' \times 18''$ | (2) |
| | | $H$ | 5.11 | 0.52 | $18'' \times 18''$ | (2) |
| | | $K_s$ | 5.15 | 0.71 | $18'' \times 18''$ | (2) |
| | WISE | W1 | 3.96 | 0.08 | Profile fit | (4) |
| | | W2 | 4.66 | 0.09 | Profile fit | (4) |
| | | W3 | 54.2 | 0.8 | Profile fit | (4) |
| | | W4 | 303.0 | 5.86 | Profile fit | (4) |
| | IRAS | IRAS2 | 509.0 | 14.3 | $>0\rlap.{'}8 \times >0\rlap.{'}8^a$ | (6) |
| | | IRAS3 | 8500 | 510.0 | $>0\rlap.{'}8 \times >0\rlap.{'}8^a$ | (6) |
| | | IRAS4 | 9980 | 798.0 | $>0\rlap.{'}8 \times >0\rlap.{'}8^a$ | (6) |
| | Herschel-PACS | Blue | 9283 | 419.0 | $35'' \times 35''$ | (11) |
| | | Green | 9241 | 415.0 | $35'' \times 35''$ | (11) |
| | | Red | 5831 | 254.0 | $35'' \times 35''$ | (11) |
| | Herschel-SPIRE | PSW | 1974 | 119.0 | $45'' \times 45''$ | (11) |
| | | PMW | 796.0 | 49.0 | $45'' \times 45''$ | (11) |
| | | PLW | 226.0 | 16.0 | $45'' \times 45''$ | (11) |
| UGC 8058 | XMM-OM | UVW1 | 1.38 | 0.01 | $4\rlap.{''}12, \times 3\rlap.{''}02$ | (12) |

Table B1
(Continued)

| Name | Instrument | Passband | S (mJy) | Error (mJy) | Integration Area ($a \times b$) | References |
|---|---|---|---|---|---|---|
| (1) | (2) | (3) | (4) | (5) | (6) | (7) |
| | | UVW2 | 0.42 | 0.01 | $2\rlap{.}''93, \times 2\rlap{.}''26$ | (12) |
| | SDSS | $u'$ | 3.66 | 0.01 | $2\rlap{.}''37 \times 2\rlap{.}''37$ | (10) |
| | | $g'$ | 11.16 | 0.01 | $2\rlap{.}''37 \times 2\rlap{.}''37$ | (10) |
| | | $r'$ | 15.7 | 0.02 | $2\rlap{.}''37 \times 2\rlap{.}''37$ | (10) |
| | | $i'$ | 9.03 | 0.03 | $2\rlap{.}''37 \times 2\rlap{.}''37$ | (10) |
| | | $z'$ | 29.11 | 0.05 | $2\rlap{.}''37 \times 2\rlap{.}''37$ | (10) |
| | 2MASS | $J$ | 58.9 | 0.01 | $36\rlap{.}''4 \times 30\rlap{.}''2$ | (2) |
| | | $H$ | 111.0 | 1.64 | $36\rlap{.}''4 \times 30\rlap{.}''2$ | (2) |
| | | $K_s$ | 191.0 | 2.83 | $36\rlap{.}''4 \times 30\rlap{.}''2$ | (2) |
| | WISE | W1 | 240.0 | 1.11 | $22'' \times 22''$ | (3) |
| | | W2 | 409.0 | 1.89 | $22'' \times 22''$ | (3) |
| | | W3 | 1310 | 6.01 | $22'' \times 22''$ | (3) |
| | | W4 | 5290 | 29.2 | $22'' \times 22''$ | (3) |
| | IRAC | IRAC3 | 605.0 | 0.10 | $5\rlap{.}''8 \times 5\rlap{.}''8$ | (8) |
| | | IRAC4 | 800.0 | 0.008 | $5\rlap{.}''8 \times 5\rlap{.}''8$ | (8) |
| | IRAS | IRAS1 | 1870 | 93.6 | $>0\rlap{.}'8 \times >0\rlap{.}'8$[a] | (6) |
| | | IRAS2 | 8662 | 93.6 | $>0\rlap{.}'8 \times >0\rlap{.}'8$[a] | (6) |
| | | IRAS3 | 20,200 | 1010 | $>0\rlap{.}'8 \times >0\rlap{.}'8$[a] | (6) |
| | | IRAS4 | 32,000 | 1600 | $>0\rlap{.}'8 \times >0\rlap{.}'8$[a] | (6) |
| | Herschel-PACS | Blue | 34,900 | 1590 | $35'' \times 35''$ | (11) |
| | | Green | 30,410 | 1370 | $35'' \times 35''$ | (11) |
| | | Red | 16,930 | 740.0 | $35'' \times 35''$ | (11) |
| | Herschel-SPIRE | PSW | 5685 | 343.0 | $50'' \times 50''$ | (11) |
| | | PMW | 2007 | 123.0 | $50'' \times 50''$ | (11) |
| | | PLW | 597.0 | 36.0 | $50'' \times 50''$ | (11) |
| IRAS 13305-1739 | GALEX | FUV | 0.048 | 0.008 | $4\rlap{.}''3 \times 4\rlap{.}''3$ | (1) |
| | | NUV | 0.071 | 0.006 | $3\rlap{.}''5 \times 3\rlap{.}''5$ | (1) |
| | SMSS | $g$ | 1.02 | 0.05 | $5\rlap{.}''28 \times 5\rlap{.}''28$ | (7) |
| | | $r$ | 1.1 | 0.06 | $5\rlap{.}''28 \times 5\rlap{.}''28$ | (7) |
| | | $i$ | 1.38 | 0.04 | $5\rlap{.}''28 \times 5\rlap{.}''28$ | (7) |
| | | $z$ | 1.1 | 0.06 | $5\rlap{.}''28 \times 5\rlap{.}''28$ | (7) |
| | 2MASS | $J$ | 3.06 | 0.17 | $14'' \times 14''$ | (2) |
| | | $H$ | 3.44 | 0.023 | $14'' \times 14''$ | (2) |
| | | $K_s$ | 3.64 | 0.02 | $14'' \times 14''$ | (2) |
| | WISE | W1 | 3.06 | 0.03 | $22'' \times 22''$ | (4) |
| | | W2 | 6.37 | 0.08 | $22'' \times 22''$ | (4) |
| | | W3 | 45.9 | 0.38 | $22'' \times 22''$ | (4) |
| | | W4 | 227.0 | 1.67 | $22'' \times 22''$ | (4) |
| | IRAS | IRAS2 | 392.0 | 17.2 | $>0\rlap{.}'8 \times >0\rlap{.}'8$[a] | (6) |
| | | IRAS3 | 1160 | 93.1 | $>0\rlap{.}'8 \times >0\rlap{.}'8$[a] | (6) |
| | | IRAS4 | 1040 | 251.0 | $>0\rlap{.}'8 \times >0\rlap{.}'8$[a] | (6) |
| | AKARI | WIDE-S | 1095 | 168.0 | $40'' \times 50''$[b] | (9) |
| | | WIDE-L | 795.3 | 79.0 | $70'' \times 90''$[b] | (9) |
| UGC 8696 | GALEX | FUV | 0.132 | 0.013 | $3\rlap{.}''5 \times 3\rlap{.}''5$ | (1) |
| | | NUV | 0.364 | 0.013 | $3\rlap{.}''5 \times 3\rlap{.}''5$ | (1) |
| | SDSS | $u'$ | 0.871 | 0.007 | $2\rlap{.}''37 \times 2\rlap{.}''37$ | (10) |
| | | $g'$ | 3.70 | 0.007 | $9\rlap{.}''30 \times 9\rlap{.}''30$ | (10) |
| | | $r'$ | 6.92 | 0.01 | $9\rlap{.}''30 \times 9\rlap{.}''30$ | (10) |
| | | $i'$ | 10.57 | 0.02 | $9\rlap{.}''30 \times 9\rlap{.}''30$ | (10) |
| | | $z'$ | 11.70 | 0.03 | $9\rlap{.}''30 \times 9\rlap{.}''30$ | (10) |
| | 2MASS | $J$ | 19.5 | 0.40 | $29'' \times 16.8''$ | (2) |
| | | $H$ | 25.3 | 0.68 | $29'' \times 16.8''$ | (2) |
| | | $K_s$ | 28.0 | 0.84 | $29'' \times 16.8''$ | (2) |
| | WISE | W1 | 23.6 | 2.17 | $15'' \times 85''$ | (3) |
| | | W2 | 36.6 | 3.37 | $15'' \times 85''$ | (3) |
| | | W3 | 232.0 | 21.4 | $15'' \times 85''$ | (3) |
| | | W4 | 1690 | 155.0 | $15'' \times 85''$ | (3) |
| | IRAC | IRAC1 | 27.0 | 2.48 | $15'' \times 85''$ | (8) |
| | | IRAC2 | 37.9 | 3.49 | $15'' \times 85''$ | (8) |
| | | IRAC3 | 79.4 | 7.31 | $15'' \times 85''$ | (8) |
| | | IRAC4 | 183.0 | 16.9 | $15'' \times 85''$ | (8) |
| | IRAS | IRAS1 | 240.0 | 17.0 | $>0\rlap{.}'8 \times >0\rlap{.}'8$[a] | (6) |

Table B1
(Continued)

| Name | Instrument | Passband | S (mJy) | Error (mJy) | Integration Area ($a \times b$) | References |
|---|---|---|---|---|---|---|
| (1) | (2) | (3) | (4) | (5) | (6) | (7) |
| | | IRAS2 | 2280 | 9.41 | $>0\rlap{.}''8 \times >0\rlap{.}''8$[a] | (6) |
| | | IRAS3 | 22,500 | 42.0 | $>0\rlap{.}''8 \times >0\rlap{.}''8$[a] | (6) |
| | | IRAS4 | 22,500 | 70.0 | $>0\rlap{.}''8 \times >0\rlap{.}''8$[a] | (6) |
| | Herschel-PACS | Blue | 24,460 | 1110 | $35'' \times 35''$ | (11) |
| | | Green | 22,320 | 1000 | $35'' \times 35''$ | (11) |
| | | Red | 12,740 | 560.0 | $35'' \times 35''$ | (11) |
| | Herschel-SPIRE | PSW | 4177 | 252.0 | $55'' \times 55''$ | (11) |
| | | PMW | 1531 | 94.0 | $55'' \times 55''$ | (11) |
| | | PLW | 427.0 | 26.0 | $55'' \times 55''$ | (11) |
| IR 14348-1447 | GALEX | FUV | 0.065 | 0.002 | $3\rlap{.}''5 \times 3\rlap{.}''5$ | (1) |
| | | NUV | 0.138 | 0.002 | $3\rlap{.}''5 \times 3\rlap{.}''5$ | (1) |
| | SMSS | $g$ | .486 | 0.87 | $3\rlap{.}''5 \times 3\rlap{.}''5$ | (1) |
| | | $r$ | 1.35 | 0.23 | $3\rlap{.}''5 \times 3\rlap{.}''5$ | (1) |
| | | $i$ | 2.31 | 0.32 | $3\rlap{.}''5 \times 3\rlap{.}''5$ | (1) |
| | | $z$ | 1.92 | 0.10 | $3\rlap{.}''5 \times 3\rlap{.}''5$ | (1) |
| | 2MASS | $H$ | 6.47 | 0.66 | $26\rlap{.}''86 \times 26\rlap{.}''86$ | (2) |
| | | $K_s$ | 9.67 | 0.92 | $26\rlap{.}''86 \times 26\rlap{.}''86$ | (2) |
| | WISE | W1 | 4.00 | .029 | $8\rlap{.}''25 \times 8\rlap{.}''25$ | (3) |
| | | W2 | 4.86 | 0.031 | $8\rlap{.}''25 \times 8\rlap{.}''25$ | (3) |
| | | W3 | 44.4 | 0.28 | $8\rlap{.}''25 \times 8\rlap{.}''25$ | (3) |
| | | W4 | 337.0 | 2.17 | $16\rlap{.}''5 \times 16\rlap{.}''5$ | (3) |
| | IRAS | IRAS2 | 560.0 | 84.0 | $>0\rlap{.}''8 \times >0\rlap{.}''8$[a] | (6) |
| | | IRAS3 | 6870 | 412.0 | $>0\rlap{.}''8 \times >0\rlap{.}''8$[a] | (6) |
| | | IRAS4 | 7490 | 674.0 | $>0\rlap{.}''8 \times >0\rlap{.}''8$[a] | (7) |
| | Herschel-PACS | Blue | 8260 | 373.0 | $35'' \times 35''$ | (11) |
| | | Green | 8385 | 376.0 | $35'' \times 35''$ | (11) |
| | | Red | 5475 | 239.0 | $35'' \times 35''$ | (11) |
| | Herschel-SPIRE | PSW | 1967 | 117.0 | $45'' \times 45''$ | (11) |
| | | PMW | 728.0 | 44.0 | $45'' \times 45''$ | (11) |
| | | PLW | 217.0 | 14.0 | $45'' \times 45''$ | (11) |
| IRAS 14394+5332 | SDSS | $u'$ | 0.187 | 0.003 | $3\rlap{.}''34 \times 3\rlap{.}''34$ | (10) |
| | | $g'$ | 0.547 | 0.002 | $3\rlap{.}''34 \times 3\rlap{.}''34$ | (10) |
| | | $r'$ | 0.945 | 0.003 | $3\rlap{.}''34 \times 3\rlap{.}''34$ | (10) |
| | | $i'$ | 1.840 | 0.006 | $3\rlap{.}''34 \times 3\rlap{.}''34$ | (10) |
| | | $z'$ | 1.870 | 0.012 | $3\rlap{.}''34 \times 3\rlap{.}''34$ | (10) |
| | 2MASS | $J$ | 2.66 | 0.18 | $14'' \times 14''$ | (2) |
| | | $H$ | 3.77 | 0.22 | $14'' \times 14''$ | (2) |
| | | $K_s$ | 3.64 | 0.31 | $14'' \times 14''$ | (2) |
| | IRAS | IRAS2 | 346.0 | 3.6 | $>0\rlap{.}''8 \times >0\rlap{.}''8$[a] | (6) |
| | | IRAS3 | 1950 | 78.0 | $>0\rlap{.}''8 \times >0\rlap{.}''8$[a] | (6) |
| | | IRAS4 | 2400 | 120.0 | $>0\rlap{.}''8 \times >0\rlap{.}''8$[a] | (6) |
| | AKARI | WIDE-S | 1780 | 130.0 | $40'' \times 50''$[b] | (9) |
| | | WIDE-L | 1770 | 410.0 | $70'' \times 90''$[b] | (9) |
| | | N160 | 910.0 | 213.0 | $70'' \times 90''$[b] | (9) |
| IRAS 17179+5444 | GALEX | NUV | 0.018 | 0.001 | $3\rlap{.}''5 \times 3\rlap{.}''5$ | (1) |
| | SDSS | $u'$ | 0.08 | 0.003 | $3\rlap{.}''84 \times 3\rlap{.}''84$ | (10) |
| | | $g'$ | 0.36 | 0.002 | $3\rlap{.}''84 \times 3\rlap{.}''84$ | (10) |
| | | $r'$ | 0.87 | 0.003 | $3\rlap{.}''84 \times 3\rlap{.}''84$ | (10) |
| | | $i'$ | 1.47 | 0.004 | $3\rlap{.}''84 \times 3\rlap{.}''84$ | (10) |
| | | $z'$ | 1.83 | 0.01 | $3\rlap{.}''84 \times 3\rlap{.}''84$ | (10) |
| | 2MASS | $J$ | 2.04 | 0.17 | $14'' \times 14''$ | (2) |
| | | $H$ | 2.09 | 0.28 | $14'' \times 14''$ | (2) |
| | | $K_s$ | 2.77 | 0.28 | $14'' \times 14''$ | (2) |
| | Spitzer | IRAC1 | 2.39 | 0.02 | $5\rlap{.}''8 \times 5\rlap{.}''8$ | (8) |
| | | IRAC2 | 3.86 | 0.02 | $5\rlap{.}''8 \times 5\rlap{.}''8$ | (8) |
| | | IRAC3 | 7.11 | 0.08 | $5\rlap{.}''8 \times 5\rlap{.}''8$ | (8) |
| | | IRAC4 | 18.3 | 0.1 | $5\rlap{.}''8 \times 5\rlap{.}''8$ | (8) |
| | IRAS | IRAS2 | 202 | 7.62 | $>0\rlap{.}''8 \times >0\rlap{.}''8$[a] | (6) |
| | | IRAS3 | 1360 | 68.1 | $>0\rlap{.}''8 \times >0\rlap{.}''8$[a] | (6) |
| | | IRAS4 | 1920 | 153.0 | $>0\rlap{.}''8 \times >0\rlap{.}''8$[a] | (6) |
| | AKARI | WIDE-S | 1340 | 40.0 | $40'' \times 50''$[b] | (9) |
| | | WIDE-L | 1380 | 89.0 | $70'' \times 90''$[b] | (9) |

Table B1
(Continued)

| Name | Instrument | Passband | S (mJy) | Error (mJy) | Integration Area ($a \times b$) | References |
|---|---|---|---|---|---|---|
| (1) | (2) | (3) | (4) | (5) | (6) | (7) |
| | | N160 | 1650 | 68.0 | $70'' \times 90''^{b}$ | (9) |
| IRAS F23529-2119 | GALEX | FUV | 0.008 | 0.001 | $3''\!.5 \times 3''\!.5$ | (1) |
| | | NUV | 0.013 | 0.001 | $3''\!.5 \times 3''\!.5$ | (1) |
| | SDSS | $u'$ | 0.032 | 0.002 | $2.73'' \times 2.73''$ | (10) |
| | | $g'$ | 0.052 | 0.001 | $2.60'' \times 2.60''$ | (10) |
| | | $r'$ | 0.105 | 0.002 | $5.10'' \times 5.10''$ | (10) |
| | | $i'$ | 0.176 | 0.003 | $4.98'' \times 4.98''$ | (10) |
| | | $z'$ | 0.232 | 0.011 | $5.74'' \times 5.74''$ | (10) |
| | 2MASS | $J$ | 0.42 | 0.03 | $4'' \times 4''$ | (2) |
| | | $K_s$ | 0.78 | 0.02 | $4'' \times 4''$ | (2) |
| | WISE | W1 | 2.0 | 0.05 | $0''\!.5 \times 0''\!.5$ | (4) |
| | | W2 | 3.11 | 0.07 | $0''\!.5 \times 0''\!.5$ | (4) |
| | | W3 | 15.3 | 0.127 | $0''\!.5 \times 0''\!.5$ | (4) |
| | | W4 | 57.6 | 0.903 | $0''\!.5 \times 0''\!.5$ | (4) |
| | Spitzer | IRAC1 | 1.9 | 0.02 | $3''\!.8 \times 3''\!.8$ | (8) |
| | | IRAC2 | 2.7 | 0.02 | $3''\!.8 \times 3''\!.8$ | (8) |
| | | IRAC3 | 4.4 | 0.07 | $3''\!.8 \times 3''\!.8$ | (8) |
| | | IRAC4 | 8.9 | 0.06 | $3''\!.8 \times 3''\!.8$ | (8) |
| | IRAS | IRAS3 | 327.0 | 55.5 | $>0'\!.8 \times >0'\!.8^{a}$ | (12) |
| | | IRAS4 | 627.0 | 132.0 | $>0'\!.8 \times >0'\!.8^{a}$ | (12) |
| IRAS 23389+0300 | GALEX | FUV | 0.004 | 0.002 | $5''\!.3 \times 5''\!.3$ | (1) |
| | | NUV | 0.009 | 0.002 | $3''\!.5 \times 3''\!.5$ | (1) |
| | SDSS | $u'$ | 0.019 | 0.002 | $2''\!.42 \times 2''\!.42$ | (10) |
| | | $g'$ | 0.053 | 0.002 | $2''\!.42 \times 2''\!.42$ | (10) |
| | | $r'$ | 0.937 | 0.003 | $2''\!.42 \times 2''\!.42$ | (10) |
| | | $i'$ | 1.82 | 0.006 | $2''\!.42 \times 2''\!.42$ | (10) |
| | | $z'$ | 1.84 | 0.01 | $2''\!.42 \times 2''\!.42$ | (10) |
| | 2MASS | $J$ | 0.77 | 0.17 | $4'' \times 4''$ | (2) |
| | | $H$ | 0.76 | 0.17 | $4'' \times 4''$ | (2) |
| | | $K_s$ | 1.22 | 0.15 | $4'' \times 4''$ | (2) |
| | WISE | W1 | 1.04 | 0.02 | $8''\!.25 \times 8''\!.25$ | (4) |
| | | W2 | 1.44 | 0.04 | $8''\!.25 \times 8''\!.25$ | (4) |
| | | W3 | 15.70 | 0.22 | $8''\!.25 \times 8''\!.25$ | (4) |
| | | W4 | 123.0 | 2.82 | $8''\!.25 \times 8''\!.25$ | (4) |
| | AKARI | N60 | 1400 | 490.0 | $40'' \times 50''^{b}$ | (12) |
| | | WIDE-S | 1320 | 396.0 | $40'' \times 50''^{b}$ | (12) |
| | | WIDE-L | 1100 | 660.0 | $70'' \times 90''^{b}$ | (12) |
| | | N160 | 2000 | 320.0 | $70'' \times 90''^{b}$ | (12) |

**Note.** Columns: (1) source name; (2) instrument; (3) passband; (4) total flux densities of the source (5) the error in the flux densities; (6) integration area used in aperture photometry in terms of length of semimajor axis $a \times$ length of semiminor axis $b$ (for circular apertures, semimajor axis=semiminor axis); (7) References. (1) Galex (https://galex.stsci.edu/GR6/?page=mastform), (2) 2MASS Extended Mission Data Release (Skrutskie et al. 2006) (3) Spoon et al. (2009), (4) Wright et al. (2010), (5) Sanders et al. (2003), (6) Wolf et al. (2018), (7) Spitzer Science Center & Infrared Science Archive (2021), (8) AKARI (Kawada et al. 2007), (9) The 16th Data Release of the SDSSs (Ahumada et al. 2020), (10) Chu et al. (2017) (11) Pollo et al. (2010).
[a] Source is flagged as resolved or marginally resolved in Sanders et al. (2003).
[b] Full width at half-maximum of the 2D Gaussian used in profile fitting photometry.

## Appendix C
## Notes on Individual Galaxies

1. *IRAS F00183-7111:* Classified as a Seyfert 2 type nucleus (Armus et al. 1989), the VLBI imaging of this galaxy reveals a GPS source at its center with a linear size equal to $\approx 1$ kpc (Norris et al. 2012). In addition, the near-IR imaging indicates a disturbed morphology and a single nucleus with large extinction values (Rigopoulou et al. 1999), while the optical morphology appears as an unremarkable smudge on a $20''$ scale (Heckman et al. 1990). This galaxy is known for its high SFR, high AGN fraction, and radio AGN-like nature. Our CIGALE modeling estimates SFR$_{IR}$ = 891.25 $M_\odot$ yr$^{-1}$, AGN fraction = $\sim$50%, and $R_{AGN} = 30$, which corroborates the above conclusions.

2. *IRAS F03538-6432:* Classified as a Seyfert 2 type nucleus (Cui et al. 2001), this galaxy is believed to be in advanced merger state because the optical morphology from Hubble Space Telescope imaging shows three distinct nuclei and a few tidal tails (Cui et al. 2001). This galaxy is a border line radio-loud galaxy with $R_{AGN} = \sim 10$.

3. *IRAS 08572+3915:* Classified as a dynamically young system formed by two disks' galaxies, which are in the process of merging (Arribas et al. 2000), this system shows the presence of a buried AGN in the northwest galaxy (Rupke & Veilleux 2013). This source presents the highest AGN fraction and the lowest SFR in our sample (see also Efstathiou et al. 2014). Moreover, this source hosts a very obscured nucleus, which is thick even in the MIR at the 9.7 $\mu$m spectral feature, and shows large water and ice opacities (Stierwalt et al. 2014). We further note that its CIAGLE estimated dust luminosity from star formation is low ($=10^{10.19}$ $L_\odot$). This is because the obscured AGN (Efstathiou et al. 2014), with a high AGN fraction = 99%, is responsible for most of the observed dust luminosity ($=10^{12.09}$ $L_\odot$; torus luminosity from CIGALE). Moreover, our radio modeling shows a very high TF = 86% for this source, which must be due to the low absorption of ionizing photons by dust (see Section 6.4).

4. *UGC 05101:* Classified as an intermediate Seyfert 1.5 type galaxy (Sanders et al. 1988) with a disturbed morphology containing a single red nucleus. It is a late-stage merger (Veilleux et al. 2002) undergoing strong circumnuclear starburst suggested by the detection of characteristic polycyclic aromatic hydrocarbon (PAH) features in MIR spectra (Martínez-Paredes et al. 2015). Although our radio SED modeling utilizes a larger dataset and frequency coverage, particularly at the lowest (74 MHz) and the highest (32.5 GHz) ends, our best-fit SED model is similar to that obtained by Clemens et al. (2010) but with two electron populations. Moreover, our derived values of $\alpha_2$ and the turnover frequency are consistent within uncertainties with Clemens et al. (2010).

5. *IRAS 10565+2448:* Classified as a H II galaxy by spectral type (Veilleux et al. 1995) undergoing minor merger (Wu et al. 1998) with a second nucleus at the northeast of the bright nucleus and a third galaxy at the tidal tail (Murphy et al. 1996). Yuan et al. (2010) classified it as a starburst-AGN composite galaxy, while Vega et al. (2008) and Vardoulaki et al. (2015) assert it to be a starburst galaxy based on the spectral index map. The best-fit radio SED model turns out to be the same as that of Clemens et al. (2010), though with a steeper $\alpha = 1.1$ and a higher turnover frequency, $\nu_t = 0.5$ GHz as compared to our analysis. The low AGN fraction = ~5% and $R_{AGN} = 1$ obtained from our analysis confirms that this galaxy is indeed starburst dominated, supporting the claim by Vardoulaki et al. (2015).

6. *IRAS 12112+0305:* Classified as a LINER-type (Kim & Sanders 1998) or a Seyfert 2 galaxy (Yuan et al. 2010), this system is in a post-merger state with two merging nuclei (Kim et al. 2002). This system shows an IR spectrum from the nuclear emission dominated by PAH emission features (Veilleux et al. 2009), hence showing no AGN signature from the IR energy diagnostics. Its radio SED is best modeled by a single-component model (C) with a turnover at 0.26 GHz. Although our turnover frequency matches within the uncertainty with that from the SYNAGE modeling (~290 MHz Nandi et al. 2021); the radio emission is due to star-forming processes with the turnover due to FFA.

7. *UGC 08058:* Classified as a Seyfert 2 galaxy (Veilleux et al. 1995) in a post-merger stage (Paspaliaris et al. 2021), this is a well-known AGN having a double-lobed radio structure at parsec scales, revealed from VLBI imaging (Ulvestad et al. 1999). Both our SED modeling and Clemens et al. (2010) SED modeling fits this galaxy with a two-component model where the fits in Clemens et al. (2010) find turnovers at 210 MHz and 3.9 GHz. Our best-fit radio SED covers more data at the low-frequency (54 MHz from LoLLS) end, and the first component of our best-fit model is without a turnover. In contrast, the turnover frequency for the second component is smaller than given in Clemens et al. (2010).

8. *IRAS 13305-1739:* Classified as a Seyfert 2 nucleus with a slight isophotal distortion in the optical image (Veilleux et al. 2002), this galaxy does not show a double-lobed radio morphology (Nandi et al. 2021). The SYNAGE modeling of this galaxy indicates a turnover frequency of ~1.7 GHz (Nandi et al. 2021), which is not corroborated by our analysis, where it is best fit with a single-PL model. This is most likely due to the inclusion of low-frequency (down to ~70 MHz from GLEAM) data in the SED modeling by us.

9. *UGC 8696:* Classified as a Seyfert 2 galaxy (Koski 1978; Sanders et al. 1988) with highly disturbed morphology with a 40 Kpc tidal tail, indicating a post-merger system. The presence of radio AGN from LOFAR imaging was proposed by Kukreti et al. (2022). Its radio SED is best modeled with a two-component model characterized by different nonthermal electron populations and turnovers (C2). The turnover frequencies are 0.16 and 7.51 GHz, respectively. Our radio SED fit model recovers a turnover at low frequency (160 MHz) because of the use of low-frequency data (54 MHz from LOLLS) by us. Clemens et al. (2010) fit a two-component model with a single turnover at 1 GHz, which is not seen in our analysis. Vega et al. (2008) note an 11% AGN fraction from radio–FUV modeling, slightly higher than our estimates.

10. *IRAS 14348-1447:* Classified as a pair of colliding galaxies (Sanders et al. 1988) with a Seyfert 1 optical nucleus (Vega et al. 2008), the system shows radio emission resulting from both radio AGN and starburst (Vardoulaki et al. 2015). Clemens et al. (2010) fit this galaxy with a turnover at 1 GHz, where our best-fit radio model turns out to be a single PL because of having more data at a lower-frequency range (<500 MHz). Vega et al. (2008) note an 8% AGN fraction from radio–FUV modeling, slightly higher than our estimates (=1.3%).

11. *IRAS 14394+5332:* Classified as a triplet with an east component having a Seyfert 2 nucleus and two close nuclei at a separation of 2.6 kpc and connected to a west component with a long tidal tail of ~35 kpc (Veilleux et al. 2002). The VLA observations conducted by Crawford et al. (1996) indicate the presence of an obscured AGN contributing the most to the radio continuum in the east component. Its radio SED is best modeled with a two-component model characterized by different nonthermal electron populations and turnovers (C2). The SYNAGE modeling of its radio SED from Nandi et al. (2021) fits a PL. This is due to the inclusion of more data, particularly at low-frequency (54 MHz from LOLLS), intermediate-frequency VLASS (3 GHz), and high-frequency VLA observations (8.4 MHz), analyzed by us (Table A1).

12. *IRAS 17179+5444:* Classified as a Seyfert 2 merger system with a single nucleus surrounded by a tidal material that seems to be extending to the southwest (Veilleux et al. 2002). This galaxy is known to host a GPS radio source, revealed by the VLBI imaging (de Vries et al. 2009). The SYNAGE modeling by Nandi et al. (2021) reveals a spectral turnover at 590 MHz, similar to that obtained by us.

13. *IRAS F23529-2119:* Classified as a Seyfert 2 galaxy (Clements et al. 1996b) and a disturbed optical morphology with an off-center nucleus leaving a broad plume to the east resulting from post-merger (Clements et al. 1996a; Rigopoulou et al. 1999). This galaxy is classified as a radio excess source ($q_{IR} \leqslant 1.4$), indicative of AGN activity (Yun et al. 2001).

14. *IRAS 23389+0300:* Classified as a Seyfert 2 nucleus and a premerger system (Rodríguez Zaurín et al. 2009), the northern galaxy hosts a double-lobed radio AGN of GPS type with a linear size of ∼830 pc (Nagar et al. 2003). Its radio SED is best modeled by the single-component model (C) with a turnover at 0.32 GHz. The SYNAGE modeling by Nandi et al. (2021) reveals a spectral turnover at 177 MHz, while our analysis recovers it at 320 MHz. This could be due to a better sampled SED used in our analysis. CIGALE estimated radio-loudness is ∼800, which supports the view that this could be a young radio AGN of GPS type (Nandi et al. 2021).

### ORCID iDs

Subhrata Dey ● https://orcid.org/0000-0002-4679-0525
Arti Goyal ● https://orcid.org/0000-0002-2224-6664
Katarzyna Małek ● https://orcid.org/0000-0003-3080-9778
Tanio Díaz-Santos ● https://orcid.org/0000-0003-0699-6083